\documentstyle[12pt,epsfig]{article}
\newcommand{\Bs}     {\mathrm{B}_{\mathrm{s}}}
\newcommand{\Bd}     {\mathrm{B}_{\mathrm{d}}}
\newcommand{\Bp}     {\mathrm{B}^{+}}
\newcommand{\Lamb}   {\mathrm{\Lambda}_{\mathrm{b}}}
\newcommand{\amp}    { {\mathcal{A}}}
\newcommand{\siga}   { \sigma_{\mathcal{A}} }
\newcommand{\abar}   { \bar{\mathcal{A}} }
\newcommand{\lik}    { {\mathcal{L}}}
\newcommand{\dlik}   { \Delta{\mathcal{L}}}
\newcommand{\dlmin}  { \Delta{\mathcal{L}}_{\mathrm{min}}}
\newcommand{\dms}    {\Delta m_{\mathrm{s}}}
\newcommand{\dmd}    {\Delta m_{\mathrm{d}}}
\newcommand{\dm}     {\Delta m}

\newcommand{\psin}   {\mathrm{ps}^{-1}}
\newcommand{\ps}     {\mathrm{ps}}
\newcommand{\CL}     {\mathrm{C.L.}}
\newcommand{\N}      {N_{\mathrm{exp}}}
\newcommand{\Nout}   {N_{\mathrm{exp}}^{\mathrm{out}}}
\newcommand{\ff}     {\mbox{\~{f}}}

\newcommand{\ie}     {{\em i.e.}}
\newcommand{\sprime} {{\boldmath{{\bf S$^\prime$}}}}
\newcommand{\spp}    {{\boldmath{{\bf S$^{\prime \prime}$}}}}

\def\PL#1#2#3{{\it Phys. Lett. }{\bf B#1 }(#2) #3}

\def\PRL#1#2#3{{\it Phys. Rev. Lett. }{\bf #1 }(#2) #3}

\def\NIM#1#2#3{{\it Nucl. Instrum. Methods A}
{\bf #1 }(#2) #3}
\def\EPJ#1#2#3{{\it Euro. Phys. J. C}
{\bf #1 }(#2) #3}

\def\Aleph{The ALEPH Coll., }
\def\Delphi{The DELPHI Coll., }
\def\L3{The L3 Coll., }
\def\Opal{The OPAL Coll., }
\def\cdf{The CDF Coll., }
\def\sld{The SLD Coll., }

\begin{document}
\begin{titlepage}
\pagestyle{empty}
\begin{center}
\Large {EUROPEAN LABORATORY FOR PARTICLE PHYSICS\\}
\end{center}
\vspace*{0.2cm}
\begin{flushright}
       CERN-EP/99-103\\
       July 15, 1999 \\
\end{flushright}

\begin{center}
\vskip 1cm
\boldmath
{\Large {\bf  The $\Bs$ oscillation amplitude analysis}}
\unboldmath
\vskip 2cm
{\large {\bf G. Boix, D. Abbaneo}}\\
\vskip 0.2cm
{\small
CERN, CH-1211, Geneva 23, Switzerland }
\end{center}
\vskip 2cm
\begin{abstract}
  The properties of the amplitude method for $\Bs$ oscillation 
analyses are studied in detail.
The world combination of measured amplitudes is converted
into a likelihood profile as a function of oscillation frequency.
A procedure is proposed to estimate the probability that the minimum observed
is due to a statistical fluctuation. This method, applied to the data
available at the time of 1999 Winter Conferences, 
gives $1 - \CL \approx 0.03$\,.
\end{abstract}

\vskip 5.cm
\begin{center}
{\it (Accepted for publication on The Journal of High Energy Physics)}
\end{center}
\end{titlepage}

\setcounter{page}{1}
\renewcommand{\thefootnote}{\arabic{footnote}}
\setcounter{footnote}{0}

\section{Introduction}
\label{sect.intro}
The first direct search for $\Bs
 - \overline{\Bs}$ oscillations was published
by ALEPH in 1994~\cite{roger}. Since then, many sophisticated analyses
have been developed by the LEP experiments, SLD and CDF~\cite{lepbosc1}. None of these 
analyses has yet been able to measure the oscillation frequency, but can exclude a range for 
the mass difference between the two $\Bs$ mass eigenstates, $\dms$,
which drives the oscillation.

In order to combine the information provided by the different analyses
in the absence of a measurement of $\dms$, a new technique~\cite{HG}, 
known as the {\em amplitude method}, was proposed. The fit
to the reconstructed proper time distribution of events tagged as mixed 
or unmixed is performed with a fixed frequency $\omega$ of the oscillating 
term, while its amplitude $\amp$ is left as the free parameter. A scan
in $\omega$ is performed and at each value the amplitude is measured. Averaging
values from different analyses is straightforward. The expected value of 
the amplitude is unity when $\omega = \dms$ (throughout this paper
$\omega$ stands for the frequency folded in the fitting function,
$\dm$ or $\dms$ indicate the frequency of the oscillations in the sample
analysed).

The range of $\omega$ for which the amplitude is found to be compatible 
with zero and incompatible with unity can be excluded. An analysis (or
a combination of analyses) has sensitivity in a given range of $\omega$ 
if the expected error on the measured amplitudes is small enough compared
to unity, so that the two values $\amp=0.$ and $\amp=1.$ can be distinguished.
In order to quote a sensitivity limit, it is normally chosen to
determine the value of $\omega$ for which a measured value $\amp=0.$ implies
that $\amp=1.$ is excluded at $95\% \ \CL$ This happens when 
$1.645\times \sigma_\amp  = 1$.

The worldwide combination, at the time of the 1999 Winter Conferences, 
of amplitude analyses shows a deviation from 
$\amp=0$ at $\omega \approx 15\, \psin$, close to the sensitivity limit of 
$14.3\, \psin$~\cite{lepbosc}, which could suggest the presence of an oscillating signal. 
The combination includes many preliminary analyses, and some new 
analyses are still expected from SLD and the LEP experiments. The significance
of the structure observed might therefore be reduced or further enhanced
in the near future as ongoing analyses are completed and published; in
any case a procedure to quantify the probability that the observed structure 
corresponds to a $\Bs$ oscillation signal is needed.
A method is proposed here, based on the generation of 
{\em toy experiments} designed to be equivalent to the world
combination of $\dms$ analyses. 

The paper is organized as follows.

In Section~\ref{sect.data} the combined amplitude at different values of
$\omega$ is presented.  The combined likelihood
profile as a function of $\omega$ is extracted from the amplitude spectrum.

In Section~\ref{sect.amp} the properties of the amplitude method are investigated.
Analytical expressions for the expected shape of the measured amplitude and its 
error are derived. The small and large frequency limits are discussed, proposing
an approximate interpretation in terms of Fourier transformations.
The probability of observing statistical fluctuations which would fake a signal in a sample
with frequency far beyond the sensitivity is also discussed.

In Section~\ref{sect.toy} the structure and the features of the toy experiment 
generator used throughout the paper are
described. A procedure to tune the parameters of the simulation in order to
reproduce the observed errors is given.

In Section~\ref{sect.sig} a procedure to extract a confidence level
value from the likelihood function is presented and discussed. The uncertainty
arising from the lack of a detailed simulation
is investigated.

\section{The experimental results}
\label{sect.data}

The combined amplitude measurements obtained from published
and preliminary analyses available at the time of the 1999 Winter 
Conferences~\cite{lepbosc}
is presented in Fig.~\ref{fig.lepbosc}.

\begin{figure}[!tb]      
  \begin{center}
    \mbox{\epsfig{figure=/afs/cern.ch/user/a/abbaneo/CORE/newbs/note/lepbosc.ps,width=130mm,%
        bbllx=0.5cm,bblly=5cm,bburx=18.5cm,bbury=22.5cm}} 
  \end{center}
  \caption[]{
    \label{fig.lepbosc}
    Current combined amplitude measurements as a function of $\omega$,
from the B Oscillation Working Group.}
\end{figure}

The observed limit is significantly smaller than the expected limit
({\it i.e.}, that which would be obtained if $\dms$ were infinitely large), due
to positive amplitude values measured in the region close to the
sensitivity limit. 

As mentioned in the introduction, the amplitude measurements
are obtained by maximizing the likelihood $L$ of the proper time
distributions of mixed and unmixed events with the amplitude
of the oscillating term $\amp$ as the free parameter, and $\omega$ fixed
at a chosen value. Denote  $\lik = -\log L$, its expansion at 
second order around the minimum of $\lik$, $\lik_{\omega}(\abar ) $ can be approximated
by
\begin{equation}
  \lik_{\omega}(\amp) \ \simeq \ \lik_{\omega}({\abar})+\frac{1}{2} 
  {\left(\frac{\amp - \abar}{\siga} \right)}_{\omega}^2\,,
  \label{eq.parab}
\end{equation}
\noindent where $\abar$ is the measured value of the amplitude,
and $\sigma_{\amp}$ is the uncertainty on $\abar$. This approximation turns out
to be very accurate in reality, $\lik_{\omega}(\amp)$ being parabolic in a wide
range around $\abar$.

From Eq.~\ref{eq.parab} it follows that, again
for each value of $\omega$:
\begin{equation}
  \lik_{\omega}(\amp=1) \ \simeq \ \lik_{\omega}({\abar})+\frac{1}{2} 
  {\left(\frac{1 - \abar}{\siga} \right)}_{\omega}^2 \, .
  \label{eq.ampone}
\end{equation}

The oscillation vanishes for $\amp=0$ on the one hand, and it averages
to zero for $\omega \to \infty$ due to finite resolution on the other. 
Therefore, the following equality can be written
\[\lik_{\omega \to \infty}({\mathrm{any}}\,\amp)=
\lik_{ {\mathrm {any}}\ \omega}(\amp=0)\,\, (\equiv \lik_{\infty}),\]
and therefore from Eq.~\ref{eq.parab},
\begin{equation}
  \lik_\infty \
 = \lik_{\omega}({\abar})+\frac{1}{2} 
  {\left(\frac{\abar}{\siga} \right)}_{\omega}^2 \, .
  \label{eq.ampzer}
\end{equation}
If Eq.~\ref{eq.ampzer} is subtracted from Eq.~\ref{eq.ampone}, the following formula is
obtained
\begin{equation}
  \dlik(\omega) \ \equiv \ \lik_{\omega}(\amp=1) - \lik_\infty \simeq \left[
  \frac{1}{2} {\left(\frac{1 - \abar}{\siga} \right)}^2 -
  \frac{1}{2} {\left(\frac{\abar}{\siga} \right)}^2 \right]_{\omega} \, ,
\label{eq.amplik}
\end{equation}
\noindent which allows the value of $\dlik$ to be calculated, for each $\omega$, 
from the fitted amplitude and its uncertainty.
This formula was already given in Ref.~\cite{HG}. 

In the derivation which follows, the total uncertainties on the measured amplitudes are used,
which is in principle not rigorous, since Eq.~\ref{eq.parab} is valid only
for the statistical part. This fact is not considered to be a problem,
since most of the uncertainties quoted as systematic errors
in the analyses combined are of statistical nature.
Typically, for inclusive analyses,
the largest sources of uncertainty on the
amplitude measurement are the fraction $f_{\Bs}$ of $\Bs$ produced in the b quark 
hadronization, the b hadron lifetimes and their mean energy.
All these parameters
are measured, and their experimental errors are propagated as systematic 
uncertainties on $\amp$. 
In some analyses the parameters considered as sources of systematic
errors were even added as free parameters in the fit, with Gaussian constraints
coming from external measurements. The part of the error identified
as systematic was thus found out {\em a posteriori \/} by running the 
minimization with those parameters fixed. 
In addition, in the region of interest ($\omega > 10\, \psin$)
the statistical errors are dominant.
The approach adopted should therefore be adequate.

The likelihood difference $\dlik(\omega)$ obtained for the data is shown in Fig.~\ref{fig.lik_dat}. 
A good parametrization for the shape of $\dlik$
is obtained with a function $f(\omega) \propto 1/\omega^\alpha$ with
\mbox{$\alpha = 1.64$}, plus some Gaussian functions to describe the deviations.
A parabolic 
fit of the three lowest points of the plot gives a minimum for
\mbox{$\omega = 14.8\,\psin$}, with a value \mbox{$\dlmin = -2.9$}. The $\dlmin+1/2$ and $\dlmin+2$ 
levels are crossed at \mbox{$(14.3-15.3)\, \psin$} 
and \mbox{$(13.0-17.5)\, \psin$} values, respectively, giving:
\begin{center}
  \begin{tabular}{rrrrlr}
    $\omega = 14.8$ &$\pm 0.5$         &$\psin$ &  & ($\dlmin+1/2$&  interval)\, , \\ 
    $\omega = 14.8$ &$^{+2.7} _{-1.8}$ &$\psin$ &  & ($\dlmin+2$ & interval)\, . \\ 
  \end{tabular}
\end{center}
These intervals would give the $\pm 1\, \sigma$ and $\pm 2\, \sigma$
uncertainties if the likelihood profile were parabolic in a range wide enough
around the minimum.
The $\dlmin+9/2$ level is crossed on the lower side only, at 
\mbox{$\omega=12.1\,\psin$}.

\begin{figure}[!tb]      
  \begin{center}
    \mbox{\epsfig{figure=/afs/cern.ch/user/a/abbaneo/CORE/newbs/note/lik_dat.ps,width=130mm,%
        bbllx=0.5cm,bblly=2cm,bburx=19cm,bbury=26.5cm}} 
  \end{center}
  \caption[]{
    \label{fig.lik_dat}
    Likelihood as a function of $\omega$ derived from the combined 
    amplitude measurements. 
    A minimum is observed for $\omega = 14.8\,\psin$.
    The parametrization described in the text is shown in (a) and (b); 
    the parabolic fit to the three lowest points in (c).
    }
\end{figure}

As discussed in the following sections, the significance of this minimum 
cannot be extracted in an analytical way, but needs to be determined with 
toy experiments.

\section{The amplitude analysis} 
\label{sect.amp}

The true proper time distribution of mixed and unmixed B meson decays is written
as follows:
\[
  {\cal P}^0_{\mathrm u,m}(t_0) \ = \  \Gamma \exp{(-\Gamma t_0)} \ 
\frac{1 \pm \cos \dm \, t_0}{2} 
\ \equiv \ \frac{E^0(t_0) \pm f^0_{\dm}(t_0)}{2} \, ,
\]
\noindent where $f^0_{\dm}(t_0)$ contains the oscillation term. 
The plus (minus) sign holds for unmixed (mixed) events. Any
difference in the decay widths of the two mass eigenstates has been neglected.

The reconstructed proper time distributions can then be written as:
\begin{equation}
  {\cal P}_{\mathrm u,m}(t) \ = \ \int_0^\infty dt_0\ \frac{E^0(t_0) \pm f^0_{\dm}(t_0)}{2} \, 
{\cal R}(t_0,t) \  \equiv \ \frac{E(t) \pm f_{\dm}(t)}{2} \, .
\label{eq.recot}
\end{equation}
\noindent For the sake of simplicity, no time dependent selection efficiency has been considered
in the calculations.
In what follows, and throughout this paper, it is assumed that 
the {\em relative uncertainty} on the 
b hadron momentum, and  
the {\em absolute uncertainty} on the decay length are Gaussian. This approximation 
follows what typically happens in real analyses, where the uncertainty on the reconstructed
b hadron momentum is found to roughly scale with the momentum itself, while the uncertainty 
on the decay length does not. This has important consequences 
in the way the two resolution components affect the amplitude shape.

Under these assumptions, the resolution function ${\cal R}(t_0,t)$ can be written as:
\begin{eqnarray}
{\cal R}(t_0,t) & = & \int_{-\infty}^\infty dp \
\frac{1}{\sqrt{2\pi} \, \sigma_p }  \, 
\exp \left( -\, \frac{(p - p_0)^2}{2\, \sigma_p^2} \right) \, 
\frac{1}{\sqrt{2\pi} \, \sigma_l }  \, 
\exp \left( -\, \frac{( pct - p_0 c t_0)^2}{2\, ( m \sigma_l)^2} \right) \, \frac{pc}{m}
\nonumber \\
&\approx & \frac{1}{\sqrt{2\pi \, [\delta_l^2 + ( \delta_p t)^2] } }  \, 
\exp \left( -\, \frac{(t - t_0)^2}{2\,  [\delta_l^2 + ( \delta_p t)^2] } \right) \, ,
\label{eq.resolu}
\end{eqnarray}
\noindent where $\delta_l\, \equiv \, \sigma_l\, m /(p_0\, c)$,
$\delta_p \, \equiv \, \sigma_p / p_0$. The approximation is valid if $\delta_p$ is
significantly smaller than one, which is anyway required to assume Gaussian errors,
since the reconstructed momentum cannot be negative.
Furthermore, $p_0$ is not accessible in real data;
the reconstructed momentum is therefore used
in the evaluation of the error from the decay length resolution:
$\delta_l  \, \approx  \, \sigma_l\, m /(p\, c)$.

A set of parameters is chosen here for the purpose of
illustration. Resolution values of $\delta_p = 0.15$ and $\delta_l = 0.14 \, \ps$ are
used; the latter one would correspond to a monochromatic sample of $\Bs$ with $p_0 = 
32~\mbox{GeV}/c$ and $\sigma_l = 250\, \mu$m. In a real analysis the normalization
of the non--oscillating component is the total number $N$
of b decays (differences in lifetime are neglected), while the oscillation
term is multiplied by \mbox{$N \, f_{\Bs} \, (1 - 2\, \eta)$}, $f_{\Bs}$ being
the fractions of $\Bs$ in the sample and $\eta$ the global mistag rate. For an inclusive
analysis \mbox{$f_{\Bs}\, (1 - 2\, \eta)$} is typically about 0.05.
The curves obtained with these parameters,
normalization factors omitted, are shown in Fig.~\ref{fig.time}.
As the frequency increases,          the oscillation amplitude is damped because of the 
resolution. For very large frequencies only the first period can be resolved.

\begin{figure}[!tb]      
  \begin{center}
    \mbox{\epsfig{figure=/afs/cern.ch/user/a/abbaneo/CORE/newbs/note/time.ps,width=130mm,%
        bbllx=0.5cm,bblly=1.5cm,bburx=19cm,bbury=26.5cm
}}
  \end{center}
  \caption[]{
    \label{fig.time}
    Reconstructed proper time distributions for the non--oscillating component, $E(t)$, 
and the oscillating component, $f_{\dm}(t)$, at different values of $\dm$. 
Resolutions of $\delta_p\,=\,0.15$ and $\delta_l\,=\,0.14\, \ps$ are assumed.
    }
\end{figure}

The fitting technique commonly used in the amplitude analysis is a simultaneous
maximum--likelihood fit to the proper time distributions of mixed and unmixed events.
Alternatively, the difference of the two distributions, {\it i.e.} the oscillating term,
can be fit with a binned $\chi^2$ method.
The two methods are discussed in the following.

\paragraph{The maximum likelihood fit.}
Using the aforementioned formalism, the likelihood function can be written as:
\begin{eqnarray*}
- \log L \ = & {\displaystyle \frac{1}{2}\int_{-\infty}^{\infty} dt \ } & 
\left[ E(t) + f_{\dm}(t) \right] \, \log \left[ E(t) + \amp f_{\omega}(t) \right] \\& + & 
\left[ E(t) - f_{\dm}(t) \right] \, \log \left[ E(t) - \amp f_{\omega}(t) \right] \ + \ 
{\mathrm{Const}} \, ,
\end{eqnarray*}
\noindent where again $\dm$ is the frequency of the oscillations in the sample analysed,
and $\omega$ is the value chosen in the fitting function.
The minimization with respect to $\amp$ leads to the condition
\begin{equation}
\int_{-\infty}^{\infty} dt \ \frac{f_{\omega}(t) \, f_{\dm}(t)  - \amp f_{\omega}^2(t)  } 
{ E(t)\, \left( 1- {\amp}^2 \frac{f_{\omega}^2(t)}{E^2(t)} \right)} \ = \ 0 \, ,
\label{eq.lik}
\end{equation}
\noindent which allows $\amp$ to be determined.

\paragraph{The {\boldmath{$\chi^2$}} fit.}
Similarly, the $\chi^2$ can be written as
\[
\chi^2 \ = \ \int_{-\infty}^{\infty} dt \ \frac{
{\left[  f_{\dm}(t) - \amp  f_{\omega}(t) \right]}^2 
}{E(t)} \, ,
\]
\noindent the minimization of which gives
\begin{equation}
\int_{-\infty}^{\infty} dt \ \frac{f_{\omega}(t) \, f_{\dm}(t)  - \amp f_{\omega}^2(t)  } 
{ E(t)} \ = \ 0.
\label{eq.chiq}
\end{equation}

Equations~\ref{eq.lik} and~\ref{eq.chiq} both give $\amp =1$ for $\omega = \dm$.
For $\omega \neq \dm$ they are equivalent if $\amp f_{\omega}(t)$ is negligible
compared to $E(t)$. 

The expression of $\amp_{\dm}(\omega)$ can be derived from Eq.~\ref{eq.chiq} as
\begin{equation}
  \amp_{\dm}(\omega) \ = \ \frac{ {\displaystyle
\int_{-\infty}^{\infty} dt \
\frac{f_{\omega}(t) \, f_{\dm}(t)}{E(t)}
}}{{\displaystyle
\int_{-\infty}^{\infty} dt \
\frac{f_{\omega}^2(t) }{E(t)} 
}} \, .
\label{eq.amp}
\end{equation}
The resulting amplitude curves for $\dm= 5,\ 10,\ 15\ \psin$ are shown in Fig.~\ref{fig.ampl}a.
On top of the curves, values obtained from the likelihood fit (Eq.~\ref{eq.lik}) are also shown.
The two fitting methods are indeed equivalent for $\omega \, \approx \, \dm$, 
as expected, while
some difference appears for $\omega \, \neq \, \dm$.

\begin{figure}[!tb]      
  \begin{center}
    \mbox{\epsfig{figure=/afs/cern.ch/user/a/abbaneo/CORE/newbs/note/all_p.ps,width=135mm,%
        bbllx=0.5cm,bblly=2cm,bburx=19cm,bbury=23.5cm}} 
  \end{center}
  \caption[]{
    \label{fig.ampl}
    (a) Expected amplitude values for $\dm = 5,\,  10,\,  15,\, \psin$. The curves refer to the $\chi^2$
minimization, the points to the likelihood fit. \\
(b) Amplitude significance curves ($\chi^2$ fit). The expected shape of $\sigma (\omega)$
is also shown. \\
(c) Expected shape of the likelihood, derived from the amplitude and its error. The dashed
line corresponds to the the limit $\dm \to \infty$.\\
Resolutions of $\delta_p\,=\,0.15$ and $\sigma_l\,=\,250\,\mu m$
are assumed for the 3 plots.
    }
\end{figure}

The expected amplitude is unity at  $\omega \, = \,  \dm$. For $\omega \, > \,  \dm$
the behaviour depends on $\dm$ (for given resolutions). In this example, for
$\dm \, = \, 15\, \psin$ the expected amplitude increases monotonically.

The expressions derived for the $\chi^2$ fit allow the expected error on the measured
amplitude to be also extracted,
\[
\chi^2(\amp+\sigma _{\amp}) - \chi^2(\amp) \ = \ 1 ,
\]
\noindent which in turn gives
\begin{equation}
\sigma _{\amp}(\omega) \ =  \ \frac{1}{ {\displaystyle \sqrt{\int_{-\infty}^{\infty} dt \
\frac{f_{\omega}^2(t) }{E(t)}}}} \, .
\label{eq.erramp}
\end{equation}
The significance of the measured amplitude is therefore:
\[
S_{\dm}(\omega) \ = \ \frac{{\amp}_{\dm}(\omega)}{\sigma _{\amp}^{\dm}(\omega)} \ = \
\frac{{\displaystyle
\int_{-\infty}^{\infty} dt \
\frac{f_{\omega}(t) \, f_{\dm}(t)}{E(t)}
}}{{\displaystyle \sqrt{
\int_{-\infty}^{\infty} dt \
\frac{f_{\omega}^2(t) }{E(t)} }
}} \, .
\]
\noindent This latter equation is correct only because $\amp$ and $\sigma_{\amp}$ 
are independent.

The amplitude significance curves for $\dm= 5,\, 10,\, 15\, \psin$ 
are shown in Fig.~\ref{fig.ampl}b.
The normalization of the error, in the same figure, is arbitrarily chosen to have 
$\sigma_{\amp}\, =\, 0.5$ at $\omega\, =\, 15\, \psin$.

The expected significance is maximal at $\omega \, =\, \dm$. For  $\omega \, >\, \dm$
it decreases without reaching zero in the range explored. The decrease is more smooth
for high values of $\dm$.

The expected shape of the likelihood, as calculated from the amplitude and its error
using Eq.~\ref{eq.amplik}, is shown in Fig.~\ref{fig.ampl}c.

\boldmath
\subsection{Limits for small and large $\dm$}
\unboldmath

In the limit of very small or very large $\dm$, some approximations
can be made in the formulae, which yield simplified expressions of 
easier interpretation.

\paragraph{Small {\boldmath{$\dm$}}.}
If $\delta_l \ll 1/\dm , \ 
\delta_p / \Gamma  \ll 1/\dm $, the oscillation is slow  and
marginally affected by the resolution. 
This limit holds in the case of $\Bd$ oscillations.
If the resolution effects are neglected, Eq.~\ref{eq.amp} can be rewritten as
\[
  \amp_{\dm}(\omega) \ = \ \frac{ {\displaystyle
\int_{0}^{\infty} dt \ \Gamma \exp(-\Gamma t) \, \cos \omega t \, \cos \dm t 
}}{{\displaystyle
\int_{0}^{\infty} dt \ \Gamma \exp(-\Gamma t) \, \cos^2 \omega t
}} \, ,
\]
\noindent which gives
\begin{equation}
\amp_{\dm}(\omega) \ \approx \ \frac{{\displaystyle
\frac{\Gamma^2}{\Gamma^2 + (\omega + \dm)^2}\, + 
\frac{\Gamma^2}{\Gamma^2 + (\omega - \dm)^2} 
}}{{\displaystyle 
1 + \frac{\Gamma^2}{\Gamma^2 + 4\, \omega^2}}} \, .
\label{eq.amp_final_slow}
\end{equation}
The resulting shape is shown in 
Fig.~\ref{fig.bd}. 
The dots superimposed are obtained
with toy experiments generated at the same value of the frequency,
including resolution effects (for details on the simulation see Section~\ref{sect.toy};
the parameters used in the generation are those of samples {\bf S} there defined).
The two shapes are in qualitatively good agreement.

\begin{figure}[!tb]      
  \begin{center}
    \mbox{\epsfig{figure=/afs/cern.ch/user/a/abbaneo/CORE/newbs/note/bd.ps,width=130mm,%
        bbllx=0.5cm,bblly=14cm,bburx=19cm,bbury=26.5cm}} 
  \end{center}
  \caption[]{
    \label{fig.bd}
The full curve gives the expected shape of the amplitude for a signal
at \mbox{$\dm = 0.5\, \psin$} when all resolution effects are neglected.
The dots are obtained with a toy experiment in which resolution
effects are simulated (where  $\delta_p\,=\,0.15$ and $\sigma_l\,=\,250\,\mu m$). 
The two shapes are in agreement.
    }
\end{figure}

\paragraph{Large {\boldmath{$\dm$}}.}
In this limit, which
corresponds to the regime of $\Bs$ oscillations,
the resolution effects dominate.
If $\dms \approx 15\, \psin$ and $\delta_p=0.15$, then 
$\delta_p / \Gamma \simeq 0.23 \, \ps$, which is larger than 
$1/ \dms \approx 0.07\, \ps$ and therefore implies that only events with small
proper time contribute to the sensitivity. Similarly, taking $\delta_l = 0.14\, \ps$
gives $\delta_l > 1/ \dms$, which implies a substantial
damping of the amplitude of the oscillating term due to the decay length resolution.
In this case, a useful approximation is to assume 
that the term $E(t)$
in Eq.~\ref{eq.amp} varies slowly compared to the fast oscillating
term $f_{\omega}(t)$, which is nonzero in a limited 
time range (Fig.~\ref{fig.time}), 
and take it out of 
the integral. In this way the expression can be simplified and rewritten in terms
of the Fourier transformation of the oscillating components,
\begin{equation}
  \amp_{\dm}(\omega) \ \approx \ \frac{ {\displaystyle
\int_{-\infty}^{\infty} dt \
f_{\omega}(t) \, f_{\dm}(t)
}}{{\displaystyle
\int_{-\infty}^{\infty} dt \
f_{\omega}^2(t) 
}} \ = \ \frac{{\displaystyle
\int_{-\infty}^{\infty} d\nu \
\ff_{\omega}(\nu) \, \ff_{\dm}(\nu)
}}{{\displaystyle
\int_{-\infty}^{\infty} d\nu \
{\ff_{\omega}}^2(\nu) 
}} \, .
\label{eq.amp_fast}
\end{equation}

The approximation is valid only if both $\omega$ and $\dm$ are large. The functions
${\ff_{\omega}}$ are shown in Fig.~\ref{fig.ff}a for a 
few different values of 
$\omega \geq 10\, \psin$. Fig~\ref{fig.ff}c shows the product of two of these Fourier
transformations to illustrate the behaviour of the ratio in Eq.~\ref{eq.amp_fast}.

\begin{figure}[!tb]      
  \begin{center}
    \mbox{\epsfig{figure=/afs/cern.ch/user/a/abbaneo/CORE/newbs/note/fou.ps,width=130mm,%
        bbllx=1cm,bblly=2cm,bburx=19.5cm,bbury=25cm}} 
  \end{center}
  \caption[]{
    \label{fig.ff}
    (a) Expected shapes of the Fourier spectra $\ff_{\dm}$ for different values of $\dm$.
    The spectra become broader and lower in amplitude when $\dm$ increases. \\
    (b) Detail of the spectra for high $\dm$. \\
    (c) Products of pairs of Fourier spectra. The resulting functions are peaked around the
    smallest of the two frequency values.\\
    Resolutions of $\delta_p\,=\,0.15$ and $\sigma_l\,=\,250\,\mu m$ are assumed.
    }
\end{figure}

With increasing $\dm$, the frequency spectra, ${\ff_{\omega}}$, become broader and 
smaller in amplitude. High {\it true} frequencies, $\dm$, have their spectrum
damped faster than low frequencies, and the peak at $\omega \approx \dm$ disappears for $\dm$
well beyond the sensitivity (Fig.~\ref{fig.ff}b). Due to the broadening of the spectra, the product
$\ff_{\omega_1}(\nu) \ff_{\omega_2}(\nu)$ is peaked around the smallest between $\omega_1$
and $\omega_2$ (Fig.~\ref{fig.ff}c). 
This fact implies that when a sample with oscillations at frequency $\dm$ is 
analysed with a function containing a frequency $\omega < \dm$, the measured amplitude is
dominated by the frequencies around $\omega$; therefore the shape of $\amp_{\dm}(\omega)$
for  $\omega < \dm$ resembles that of $\ff_{\dm}(\omega)$. For  $\omega > \dm$ the
frequencies around $\dm$ are always tested, with a normalization factor which increases
with $\omega$ (Eq.~\ref{eq.amp_fast}, the denominator decreases very fast), 
therefore $\amp_{\dm}(\omega)$ increases monotonically.

\vskip 0.5cm

In order to understand better the effect of the decay length and proper time resolution,
it is useful to study them separately. Setting $\delta_p = 0$ in Eq.~\ref{eq.resolu},
the following simplified expression can be obtained,
\[
\ff_{\dm}(\omega) \ = \ \frac{1}{2} \, \left[
  \frac{\Gamma^2}{\Gamma^2 + (\omega + \dm)^2}\, + 
  \frac{\Gamma^2}{\Gamma^2 + (\omega - \dm)^2} \right] \, \exp{\left( - \,
  \frac{\delta_l^2 \omega^2}{2} \right) } \, ,
\]
\noindent which shows that the decay length resolution is responsible
for the damping of the high frequencies.

Considering the momentum resolution alone, the following expression is obtained,
\[
\ff_{\dm}(\omega) \ = \ \int_{-\infty}^{\infty} d\nu \ 
\frac{1}{2} \, \left[
  \frac{\Gamma^2}{\Gamma^2 + (\nu + \dm)^2}\, + 
  \frac{\Gamma^2}{\Gamma^2 + (\nu - \dm)^2} \right] \, \exp{ \left( - \,
  \frac{(\omega - \nu)^2}{2\, (\delta_p \omega)^2} \right)}
\, ,
\]
\noindent which shows that the momentum resolution causes the broadening of the frequency 
spectrum (as intuitively expected, since a shift in the reconstructed momentum is 
equivalent to a change of scale on the time axis). 
For $\dm = 15 \, \psin$ and $\delta_p = 0.15$, the width of the frequency spectrum is 
dominated by the momentum resolution. 

A broader frequency spectrum corresponds to a broader structure in the amplitude spectrum, 
or, equivalently, to higher correlations between values of the amplitude
measured at different frequencies. This property is relevant for the confidence level
estimation as explained in Section~\ref{sect.sig}.

\subsection{Fluctuations}
\label{subsect.fluct}

  The expected shape of the likelihood for a sample
with oscillations at a frequency far beyond the
sensitivity is shown in Fig.~\ref{fig.ampl}c. In a given frequency range,
statistical fluctuations of the likelihood can produce values below 0 which
can fake a signal. The probability of observing that 
$\dlik$ is lower than a given value $ \overline{ \dlik }$ at a given frequency $\omega$
can be estimated from Eq.~\ref{eq.amplik},
using the fact that the errors on the measured amplitudes are found
to be Gaussian with high precision:
  \begin{equation}
 {\cal {P}}( \dlik, \omega  ) \ \equiv \
 {\cal {P}}( \dlik < \overline{ \dlik } )_{\omega} \ = \
  \frac{1}{2} \, \mbox{erfc}\left[  \left. \left( - \overline{\dlik}\, \siga(\omega) +
        \frac{1}{2\, \siga(\omega)} \right)  \right/ {\sqrt{2}}\,  
  \right] \, .
  \label{eq.fluct}
\end{equation}

The function $ {\cal {P}}( \dlik, \omega  ) $ is shown in Fig~\ref{fig.conto}a, where
the same parameters and normalization as for Fig.~\ref{fig.ampl} are used.
This function can be used as an estimator of the signal--ness of a given sample.
Estimator contours, equidistant on a logarithmic scale, are drawn in 
Fig.~\ref{fig.conto}b.
Small negative values are most probable at high
frequencies, while for larger negative values the maximum of the probability is found 
at lower frequencies.

\begin{figure}[!tb]      
  \begin{center}
    \mbox{\epsfig{figure=/afs/cern.ch/user/a/abbaneo/CORE/newbs/note/con_mc.ps,width=120mm,%
        bbllx=0.5cm,bblly=1cm,bburx=19cm,bbury=26.cm}} 
  \end{center}
  \caption[]{
    \label{fig.conto} 
    (a) Probability to find a $\dlik$ value lower than  $\overline{ \dlik }$ 
    at a frequency $\omega$. \\
    (b) Contours of equal probability in the ($\overline{ \dlik }, \omega$) 
plane.
    Values between $-3$ and $-5$ are found 
    with higher probability when the error on the amplitude is from 0.5 
    (at $\omega = 15\, \psin$ in this example) to 0.3 ($\omega \approx 11\, \psin$). 
    The contours shown are equidistant 
    on a logarithmic scale. \\
    Resolutions of $\delta_p\,=\,0.15$ and $\delta_l\,=\,0.14\,\ps$
    are assumed.
    }
\end{figure}

\section{The toy experiments}
\label{sect.toy}

The probability that the minimum observed in the likelihood (Fig.~\ref{fig.lik_dat})
is caused by a fluctuation can be evaluated by means of toy experiments with the above 
estimator (Eq.~\ref{eq.fluct}).

In a general case, the depth of the likelihood minimum
can be translated to a statistical significance
in the approximation that the likelihood is parabolic,
which is not the case here.

At each frequency point, the probability that the measured $\dlik$ is lower than a given value $\overline{\dlik}$ can be calculated as explained in Section~\ref{subsect.fluct},
starting from the errors on the measured amplitude. 
This procedure cannot be applied to the minimum, since $\dlmin$ is not
an ``unbiased'' value, but it is chosen as the lowest value found over a certain
frequency range explored.

 The sum of the probabilities 
of obtaining a likelihood value lower than
observed at all points where the amplitude is measured does not
provide a good estimate either. The different points are highly correlated 
and they cannot fluctuate independently, therefore the sum of the
individual probabilities would give a gross 
overestimate
of the overall probability of finding a minimum as or more unlikely than the one observed.

The only viable possibility is to calibrate the significance of the structure observed
with the help of toy experiments. 
The worldwide combination includes many analyses, and a detailed simulation of each of them 
is highly impractical. The procedure adopted here is to choose a set of parameters for
the generation of the toy 
experiments such that each experiment is equivalent to the world average. 
The set of parameters cannot be uniquely determined from the data: it turns out that
some parameters need to be fixed {\it a priori}, and therefore the dependence of the
result obtained upon the particular choice adopted needs to be understood.
The possible effects of the lack of a detailed
simulation are investigated in Section~\ref{sect.corre} by studying the dependence 
of the correlations 
in the amplitude measurements upon the parameters chosen to generate the toy experiments.

\subsection{Generation}
\label{subsect.gene}

The basic features of the toy experiments used to estimate the significance
of the likelihood minimum can be summarized as follows.

\begin{itemize}

\item Bottom hadron species are generated according to a chosen
  composition.
\item For each species, the {\it true} proper time $t_0$ of each b hadron 
  is generated according to an exponential with decay constant
  equal to its width, $\Gamma$, multiplied by a given efficiency function.
\item Neutral B mesons are allowed to mix. Mixed and unmixed particles have
  their proper time distributions modified by the appropriate
  oscillating term, with given frequency.
\item The {\it true} momentum $p_0$ is generated according 
  to a Peterson distribution, tuned to reproduce a given mean scaled energy 
  $\langle x_b \rangle $.
\item The {\it true} decay length is then obtained, for each b hadron, from
  \[
  l_0 \,=\, \frac{t_0 p_0 }{m} \, c \, .
  \]
\item A smearing is applied to the {\it true} decay length and momentum
  according to given resolution functions, to obtain the
  {\it measured} decay length and momentum, $l$ and $p$.
\item The {\it measured} proper time is hence calculated as
  \[
  t\,=\, \frac{l m}{pc}  \, .
  \]
\item A mixed/unmixed tag is assigned to the generated hadrons 
  using specified mistag rates.
\item The udsc background is neglected.
\end{itemize}

\subsection{The choice of the parameters}
\label{subsect.param}

The only information, at the level of the world combination,
which can drive the choice of the parameters for the simulation is provided
by the errors on the measured amplitudes.
The step at $\omega = 15 \, \psin$ is due to some analyses in which
the scan was not performed beyond that value of the frequency. 
The step at $\omega = 19 \, \psin$ is due
to the SLD analyses, for which no  measurement was provided for  $\omega > 19 \, \psin$.
In all what follows the four points with  $\omega > 19 \, \psin$ are ignored, 
in order to reduce the pathologies in the error shape. 

The errors on the measured amplitudes can be formally written as 
(see also Eq.~\ref{eq.erramp})
\begin{equation}
\sigma^{-1}_{\amp}(\omega) \ = \ \sqrt{N} \, f_{\Bs} \, (1-2\, \eta) \, 
\Sigma(\delta_p, \sigma_l, \omega) \, .
\label{eq.err_tun}
\end{equation}
The factor $\kappa \, = \, \sqrt{N} \, f_{\Bs} \, (1-2\, \eta)$ gives the
normalization of the error distribution, without affecting the shape, and
obviously the three parameters can not be disentangled. It is chosen to 
fix $f_{\Bs}$ and $\eta$ to some ``typical'' values (namely  $f_{\Bs}=0.15$,
 $\eta=0.25$), and adjust $N$ to fit the data errors. The effect of a different choice
which yields the same $\kappa$ value is investigated in Section~\ref{sect.corre}.

The decay length and momentum resolution terms both affect the shape of the
measured error as a function of $\omega$. The sensitivity is not enough
to get a reliable simultaneous determination of both. It is thus chosen to fix
$\delta_p$, again, to a ``typical'' value of $\delta_p = 0.15$, and tune
the value of $\sigma_l$. This choice is preferred 
because, as explained later,  $\delta_p$ plays an important r\^ole in the 
determination of the confidence level, and needs anyway to be varied over a wide range
to check the stability of the result obtained.

Samples are generated at three starting points for  $\sigma_l$, which are chosen to be
$200\, \mu$m, $250\, \mu$m and $300\, \mu$m, each with 30000 events and the other 
parameters as described above, and reported in more detail 
in Section~\ref{subsect.sample}.

For each value of $\sigma_l$, the number of events is tuned by minimizing the sum
of the differences with the data errors, 
\[
\sum_i \left( \siga^{\mathrm{data}} - \siga^{\mathrm{toy}} \right)^2_i \, ,
\]
where the scaling low of
Eq.~\ref{eq.err_tun} is used (Fig.~\ref{fig.opterr}a). The three minima found are then
compared and interpolated with a parabolic fit (Fig.~\ref{fig.opterr}b)
to find the absolute minimum, which turns out to be very close to $250\, \mu$m.
The number of events needed at this point is 34000.

\begin{figure}[!tb]      
  \begin{center}
    \mbox{\epsfig{figure=/afs/cern.ch/user/a/abbaneo/CORE/newbs/note/opt.ps,width=130mm,%
        bbllx=0.5cm,bblly=18.5cm,bburx=19.5cm,bbury=26.5cm}} 
  \end{center}
  \caption[]{
    \label{fig.opterr}
    (a) Optimization of the number of events for three different values of the
    decay length resolution. \\
    (b) Choice of the optimal decay length resolution.
    }
\end{figure}

\subsection{Samples description}
\label{subsect.sample}

On the basis of the procedure described in Section~\ref{subsect.param}, a
set of parameters {\bf S} is defined as follows:

\begin{itemize}
\item a single {\it purity} class: $15\% \, \Bs ~~38\% \, \Bd ~~38\% \, \Bp ~~9\% \, \Lamb$; 
\item a single {\it tagging} class: mistag rate $\eta\,=\,25\%$ for all species; 
\item a single {\it resolution} class:
  \[ \left. 
    \begin{array}{ll}
      \sigma_l\,=\,250\,  \mu m\\
      \sigma_p/p_0 \,=\,0.15
    \end{array}
  \right\} \mbox{ both Gaussian with no tails} \, ;
  \]
\item Monte Carlo parametrized efficiency (taken from the analysis of Ref.~\cite{Olivier}). 
  The curve is shown in Fig.~\ref{fig.eff};
\item b hadron masses and lifetimes, and $\dmd$ from Ref.~\cite{PDG};
\item   $\langle x_b \rangle = 0.7$;
\item $\dms$ fixed at different values, according to the study considered;
\item statistics of 34000 b hadron decays.
\end{itemize}

\begin{figure}[!tb]      
  \begin{center}
    \mbox{\epsfig{figure=/afs/cern.ch/user/a/abbaneo/CORE/newbs/note/eff.ps,width=130mm,%
        bbllx=0.5cm,bblly=18.5cm,bburx=19cm,bbury=26.5cm}} 
  \end{center}
  \caption[]{
    \label{fig.eff}
    Shape of the reconstruction efficiency as a function of true proper time. 
    The normalisation of the vertical scale is arbitrary.
    }
\end{figure}

A second set of parameters \sprime\ is defined to generate a second
family of toy experiments. The momentum resolution is chosen to be
$\sigma_p/p_0\,=\,0.07$, which is significantly better than
what is typically achieved in inclusive analyses. 
In order to keep the agreement with the
world average errors on the measured amplitudes, the number of events
is reduced to 29000 (obtained with the procedure described 
in Section~\ref{subsect.param}).
The other parameters are left unchanged.
These experiments are used in the following to investigate the dependence
of the confidence level upon the momentum resolution.

The errors on the amplitude, $\siga$, obtained with these two 
sets of experiments are 
compared to the errors from the combined data in Fig.~\ref{fig.error_comp}.
The step at $\omega = 15\, \psin$ could be reproduced
by averaging, for each ``experiment'', two ``analyses'',
of which one has its scan stopped at that point. No attempt was made in this direction.

\begin{figure}[!tb]      
  \begin{center}
    \mbox{\epsfig{figure=/afs/cern.ch/user/a/abbaneo/CORE/newbs/note/error_comp.ps,width=130mm,%
        bbllx=0.5cm,bblly=14cm,bburx=19cm,bbury=26.5cm
        }}
  \end{center}
  \caption[]{
    \label{fig.error_comp}
    Amplitude errors comparison: simulated experiments {\it versus} world average data.
    }
\end{figure}

A third set of samples \spp\ with $\delta_p= 0.15$, $\sigma_l = 200 \, \mu$m and
statistics of 16500 decays (which correspond to the optimization of 
Fig.~\ref{fig.opterr}) is used to investigate the dependence of the 
confidence level upon the decay length resolution.

Finally samples of type {\bf s} are defined from samples {\bf S} by increasing $f_{\Bs}$
by a factor of five (\ie\ having $f_{\Bs} = 0.75$) and reducing the statistics by a factor of
25 (which gives 1360 decays).

In Fig.~\ref{fig.nosig} the expected shape of the amplitude and the likelihood is shown,
as obtained by averaging 1000 samples of type {\bf S}, generated with $\dms = 150\, \psin$.
The expected value is consistently zero, and the errors on $\amp$ are Gaussian, which confirms 
the validity of the amplitude method to set limits on the oscillation frequency.

\begin{figure}[!tb]      
  \begin{center}
    \mbox{\epsfig{figure=/afs/cern.ch/user/a/abbaneo/CORE/newbs/note/nosig.ps,width=130mm,%
        bbllx=0.5cm,bblly=8.5cm,bburx=19cm,bbury=23cm}} 
  \end{center}
  \caption[]{
    \label{fig.nosig}
    (a) Expected amplitude and error for samples of type {\bf S}, with $\dms = 15 \psin$, as a function
    of $\omega$. \\ (b) Expected likelihood shape. The plots are obtained by averaging 2000 samples.
    Resolutions of $\delta_p\,=\,0.15$ and $\sigma_l\,=\,250\,\mu m$ are assumed.
    }
\end{figure}

\section{The estimate of the significance}
\label{sect.sig}

As demonstrated in Section~\ref{subsect.fluct}, the probability that, at a given point in the 
frequency scan, a value of the likelihood $\dlik < \overline{ \dlik }$ be found,
can be calculated, given 
$\overline{ \dlik }$, from the error on the measured amplitude, which is available from the data.

For the purpose of establishing the significance of the minimum, however, this probability
is not enough, 
since what is needed is the probability that anywhere in the range explored a configuration
more unlikely than the one observed may appear (in the hypothesis of large $\dms$).
This significance is driven not only by the errors, but also by the correlations between the 
amplitude measurements at different frequencies, which are not controlled from the data,
and might depend on the particular combination of parameters chosen for the simulation. 
It is therefore mandatory to identify the most relevant sources of systematic uncertainty which
might affect the extraction of the confidence level.
This point is investigated in the next section.

\subsection{Correlations}
\label{sect.corre}

From the discussion of Section~\ref{sect.amp}, it turns out that the momentum resolution
is the most critical parameter to determine the point--to--point correlation in the
amplitude scan.
In a sample with better momentum resolution,
correlations are smaller and therefore the probability of having significant
deviations from $\amp = 0$ in a sample with no signal is larger, 
in a given frequency range explored.

In order to investigate the dependence of the point--to--point correlation
upon the parameters used in the generation, a sensitive quantity is the average
difference between amplitudes measured at two given points
in the frequency scan. If there were no correlations, this difference could be written in terms
of the errors on the amplitude as
\[
\langle \, \mid \amp_i - \amp_j \mid \, \rangle  \ =\  \sqrt{\frac{2}{\pi}} \, 
\sqrt{ {\sigma_{\amp}^i}^2 + {\sigma_{\amp}^j}^2 } \, .
\]
Correlations reduce this value if $i$ and $j$ are close enough.
A scan in steps of $0.25\, \psin$ is assumed, as for the data analyses. 

For each of the four set of parameters, {\bf S}, \sprime, \spp\ and {\bf s} defined
in Section~\ref{subsect.sample}, 150 samples are produced, and the quantity 
$\langle \, \mid \amp_i - \amp_j \mid \, \rangle$ is calculated, for $i-j \, = \, 1,\ 4,\ 7,\ 10$.
The results are shown in Fig.~\ref{fig.deriv}, where they are compared with the expectation
for no point--to--point correlation.

\begin{figure}[!tb]      
  \begin{center}
    \mbox{\epsfig{figure=/afs/cern.ch/user/a/abbaneo/CORE/newbs/note/der_comp_lab.ps,width=130mm,%
        bbllx=0.5cm,bblly=1.5cm,bburx=19cm,bbury=26.cm}} 
  \end{center}
  \caption[]{
    \label{fig.deriv}
    Point--to--point fluctuations for four sets of samples. From top to bottom,
    the average difference between points distant 1--4--7--10 steps in the amplitude
    scan are shown. The lines correspond to the limit of no correlation between points.
    }
\end{figure}

Compared to the most ``realistic'' samples, {\bf S}, the largest deviation is observed,
as expected, when the momentum resolution is changed (samples \sprime).
At low $\omega$, the difference between the no-correlation
limit (curve) and the values found in the simulation (markers), 
decreases rapidly as the distance between the points increases:
for $i-j\, = \, 4$ ($\Leftrightarrow \, \Delta \omega = 1\, \psin$) it is reduced by about a factor
of two compared to $i-j\, = \, 1$, so $\Delta \omega = 1\, \psin$ can be taken as an estimate of the
``correlation length'' at small frequencies. When $\omega$ increases, the difference between the 
curve and the simulation remains substantial even when the points are a few $\psin$ apart, 
demonstrating the increase of the correlation length with $\omega$.

Samples \sprime\ can be used to estimate a ``systematic uncertainty'' 
on the confidence level obtained,
coming from the specific choice of the parameters used in the simulation.

\subsection{The Confidence Level}

The significance of the minimum observed in the $\dlik$ distribution 
(Fig.~\ref{fig.lik_dat}) is estimated by computing the probability
that a structure as or more unlikely is
observed in a sample with $\dms$ far beyond the sensitivity.

In order to do that, it is taken into account that the
probability of observing a given value of $\dlik$ is a non--trivial function of $\omega$.
Probability contours in the ($\dlik, \omega$) plane (as in Fig.~\ref{fig.conto}b) are
built from the data errors. The contour corresponding to the data sample is computed.
$\N \, = \, 2000$ samples of type {\bf S} with 
$\dms = 150\, \psin$
are analysed and the number $\Nout$ of those that give a minimum $\dlmin\,<\,0$ outside 
the contour corresponding to the data is recorded. Since the expected value of the likelihood 
is positive for all frequencies (see Fig.~\ref{fig.nosig}b), occasionally the minimum in the range 
$0-19\, \psin$ is also positive. These minima are not counted, independently of the frequency 
at which they occur, since thay can not be interpreted as a signal of oscillations.

The population of the toy experiments in the  ($\dlik, \omega$) plane along with the point
corresponding to the data sample, is shown in Fig.~\ref{fig.cl15}.

\begin{figure}[!tb]      
  \begin{center}
    \mbox{\epsfig{figure=/afs/cern.ch/user/a/abbaneo/CORE/newbs/note/cont_cl_15.ps,width=130mm,%
        bbllx=1cm,bblly=14cm,bburx=19cm,bbury=26.5cm}} 
  \end{center}
  \caption[]{
    \label{fig.cl15}
    Minima of  $\dlik$ for 2000 samples of type {\bf S}, with $\dms = 150\, \psin$.
    The curves represent contours of equal probability of observing a value of $\dlik$ 
    smaller than $ \overline{ \dlik }$, as a function of $\omega$ 
    (as in Fig.~\ref{fig.conto}).
    }
\end{figure}

The confidence level is computed as
\begin{equation}
1 - \CL \ \equiv \ \frac{\Nout}{\N} \ = \ 0.021 \pm 0.003 \,.
\label{eq.cl15}
\end{equation}

The study is repeated with 2000 samples of type \sprime, and yields
\begin{equation}
1 - \CL \ = \ 0.033 \pm 0.004 \,.
\label{eq.cl7}
\end{equation}

This value has to be understood as a conservative estimate of the probability of 
statistical fluctuations,
since it is obtained with experiments built to have lower point--to--point correlations
than that expected for the average of real analyses. The distribution of the minima 
for this case is shown in Fig.~\ref{fig.cl7}.
The difference between the values of Eq~\ref{eq.cl15} and Eq.~\ref{eq.cl7}
gives an upper limit for the uncertainty coming from the
lack of a detailed simulation. 

\begin{figure}[!tb]      
  \begin{center}
    \mbox{\epsfig{figure=/afs/cern.ch/user/a/abbaneo/CORE/newbs/note/cont_cl_7.ps,width=130mm,%
        bbllx=1cm,bblly=14cm,bburx=19cm,bbury=26.5cm}} 
  \end{center}
  \caption[]{
    \label{fig.cl7}
    Minima of  $\dlik$ for 2000 samples of type \sprime\, with $\dms = 150\, \psin$.
    The same curves as in Fig~\ref{fig.cl15} are shown.
    }
\end{figure}

The probability that the current result of the world combination of $\Bs$
oscillation analyses is due to a statistical fluctuation can be therefore quantified to
be around $3\%$. The uncertainty on this number coming from the inaccuracies
of the simulation is below $1\%$.

\subsection{Comparison with the oscillation hypothesis}

In order to check that the amplitude spectrum
observed in the data is in qualitative agreement with the hypothesis 
of oscillations, 500 samples of type {\bf S} have been produced, 
with $\dms\, = \, 14.8\, \psin$. The expected amplitude and error at each
frequency value are shown in Fig.~\ref{fig.ave148}, with the data points 
superimposed. The agreement is good over the whole frequency range.

\begin{figure}[!tb]      
  \begin{center}
    \mbox{\epsfig{figure=/afs/cern.ch/user/a/abbaneo/CORE/newbs/note/sig148.ps,width=130mm,%
        bbllx=1cm,bblly=15cm,bburx=19cm,bbury=24.5cm}} 
  \end{center}
  \caption[]{
    \label{fig.ave148}
    Average amplitude and expected error as a function of $\omega$ for a signal at 
    \mbox{$\dms\, = \, 14.8 \, \psin$}. The amplitude values, obtained by averaging 500 toy 
    experiments, are in good agreement with the data measurements (solid points).
    }
\end{figure}

A quantitative study of the compatibility of the
data with the signal hypothesis would require to perform a fine 
scan on $\dms$ with many samples at each value, in order to define a probability that the results
observed are produced by an oscillation with a frequency in the range explored.
This kind of study is not attempted here.

A simple check is performed instead. The 500 samples with oscillation at a value
\mbox{$\dms\,=\,14.8\,\psin$} are analysed in terms of their incompatibility with the 
no--oscillation hypothesis. The scatter plot of the likelihood minima in the 
($\overline{\dlik}, \omega$) plane, as for the samples with \mbox{$\dms\,=\,150\,\psin$}, is presented 
in Fig~\ref{fig.data_cl}.

\begin{figure}[!tb]      
  \begin{center}
    \mbox{\epsfig{figure=/afs/cern.ch/user/a/abbaneo/CORE/newbs/note/plots/cont_cl_15_sig.ps,width=130mm,%
        bbllx=0.5cm,bblly=15.cm,bburx=19.5cm,bbury=26.5cm}} 
  \end{center}
  \caption[]{
    \label{fig.data_cl}
    Minima of  $\dlik$ for 500 samples of type {\bf S}, with $\dms = 14.8\, \psin$.
    The same curves as in Fig~\ref{fig.cl15} are shown.
    }
\end{figure}

An enhanced density in the region \mbox{$14\,\psin<\omega<16\,\psin$}, 
\mbox{$-3<\dlik_{\hbox{min}}<-1$} is shown in the plot. A cluster of experiments with minima at 
\mbox{$\omega\,=\,19\,\psin$} is also
clearly visible: for these experiments the lowest point of the likelihood was at the
boundary of the region analysed.
Experiments with $\dlik_{\hbox{min}}<-5$ appear at frequencies lower than the true one, where
fluctuations which can produce deep minima are more likely.

Out of these 500 samples, 80 were found outside the estimator contour corresponding to the
data, which gives a probability of $16 \%$. If the data results were perfectly
``typical'' compared to the toy samples, the expected result
would be $50 \%$.

\section{Conclusion}

The likelihood profile as a function of $\omega$ derived from the combined
amplitude measurements available at the time of the 1999 Winter Conferences
shows a minimum at $\omega \, = \, 14.8 \, \psin$.
The depth of the minimum compared to the asymptotic value for $\omega \to \infty$
is $\dlmin \, = \, -2.9$. The intervals at which the $\dlmin+1/2$ and $\dlmin+2$
levels are crossed are:
\begin{center}
  \begin{tabular}{rrrrlr}
    $\omega = 14.8$ &$\pm 0.5$         &$\psin$ &  & ($\dlmin+1/2$&  interval)\, , \\ 
    $\omega = 14.8$ &$^{+2.7} _{-1.8}$ &$\psin$ &  & ($\dlmin+2$ & interval)\, . \\ 
  \end{tabular}
\end{center}

The significance of the minimum observed in the likelihood cannot be calculated
analytically, but needs to be calibrated using toy experiments. 
With the method proposed here, the 
probability that the result observed is produced by a statistical fluctuation
in a sample with no signal is found to be:
\[
1- \CL \ \approx \ 3\, \% \, .
\]

The uncertainty on this estimate coming from the lack of a detailed simulation
of the individual analyses contributing to the average is estimated to be below $1\, \%$.

\section{Acknowledgements}

We would like to thank Hans--Guenther Moser for following our work and giving
constructive feedback, and Olivier Leroy for many lively discussions and for checking
all calculations throughout the paper. We are grateful to Patrick Janot and
Gigi Rolandi for their advice and help all along the development of the
analysis and their careful reading of the manuscript.


\newpage
\pagestyle{empty}

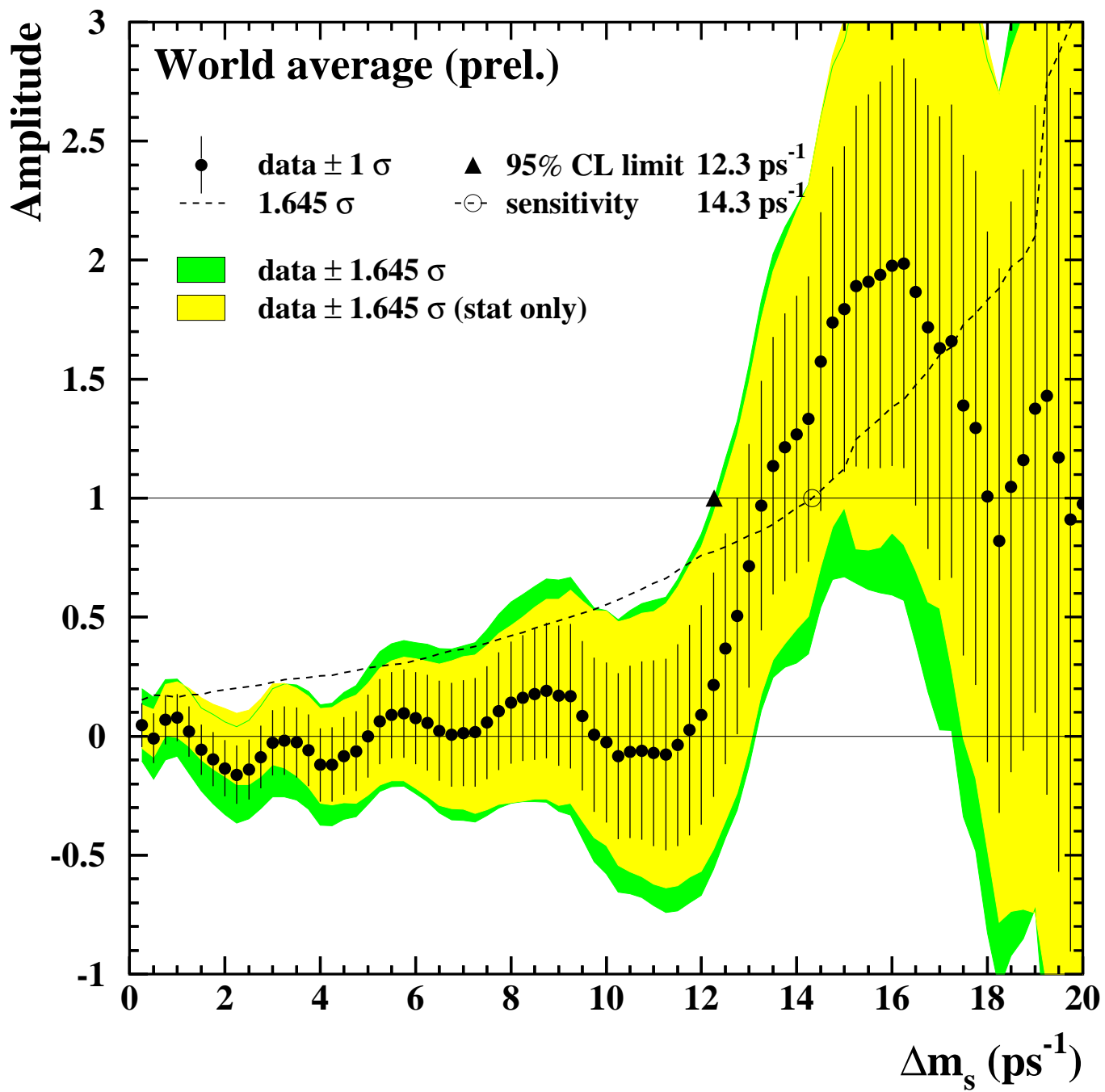
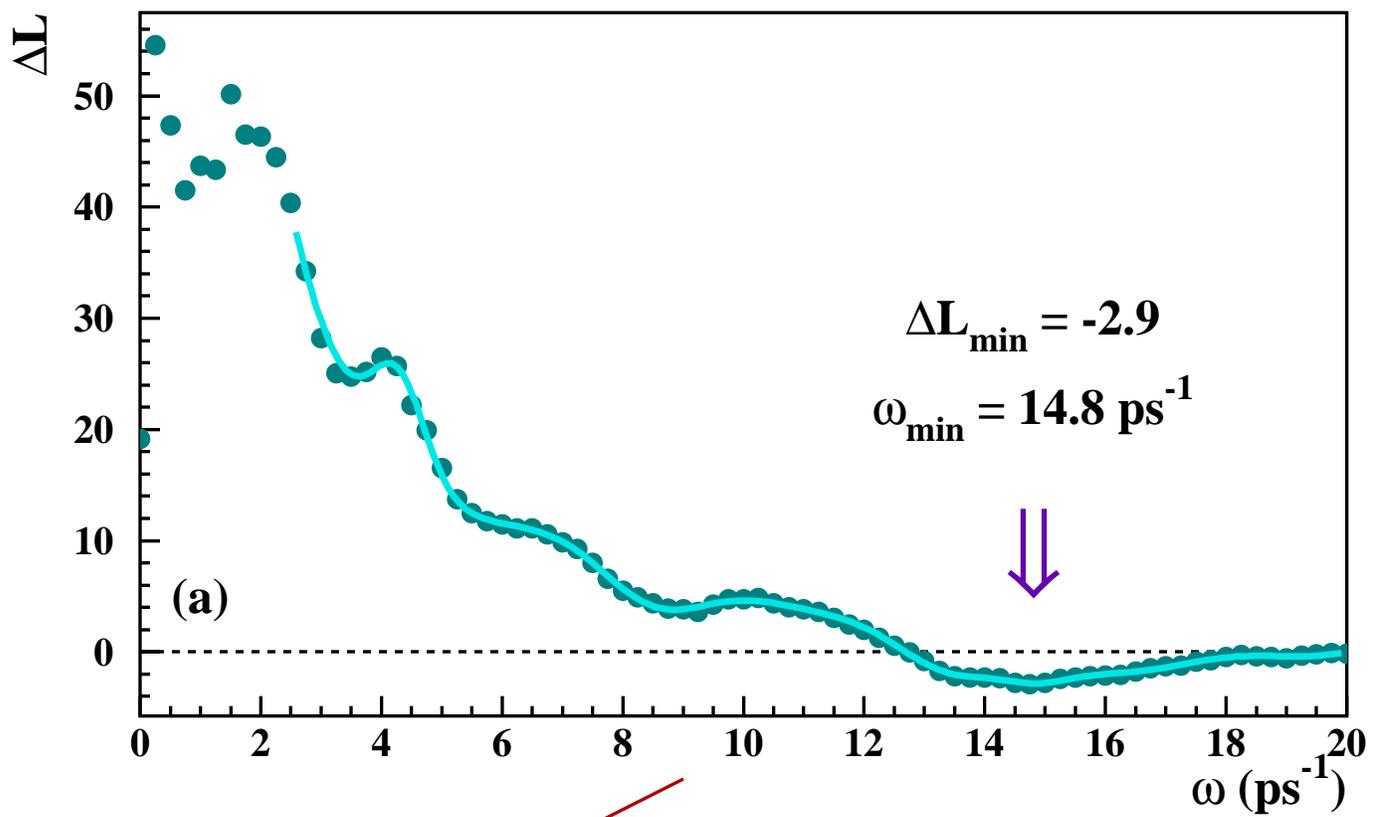
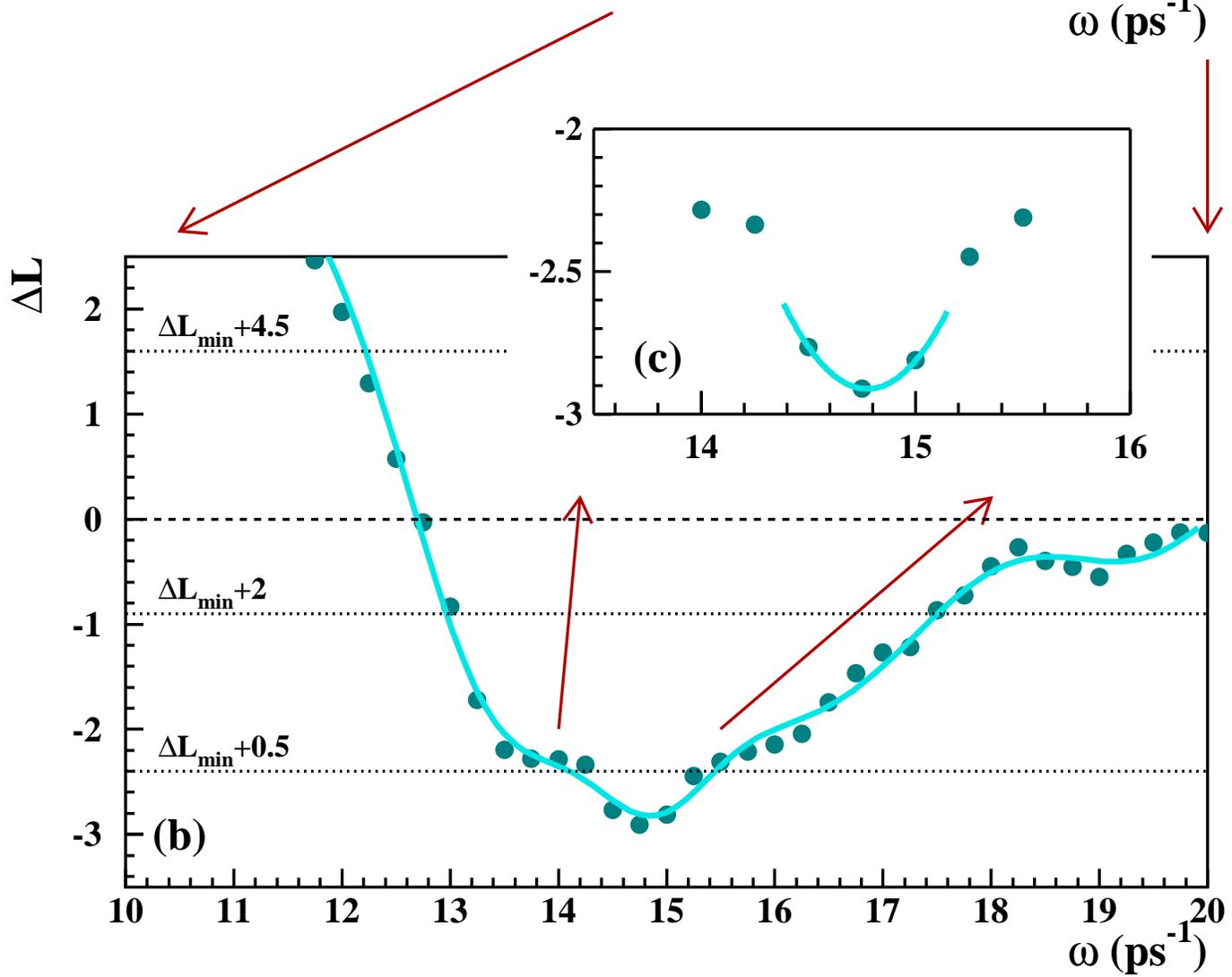
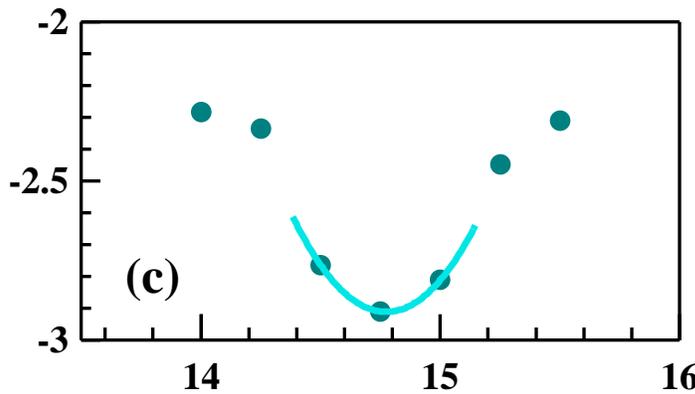
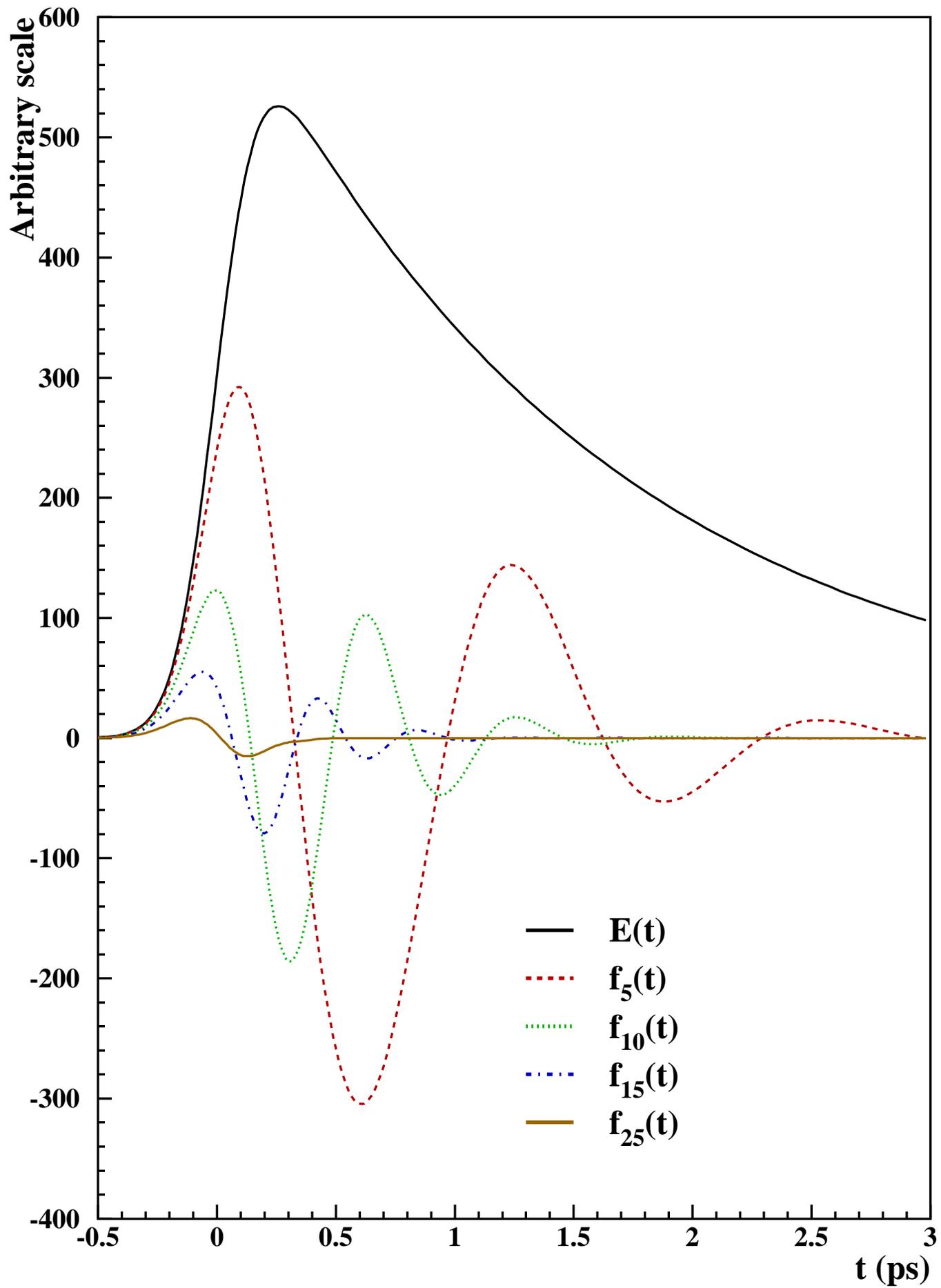
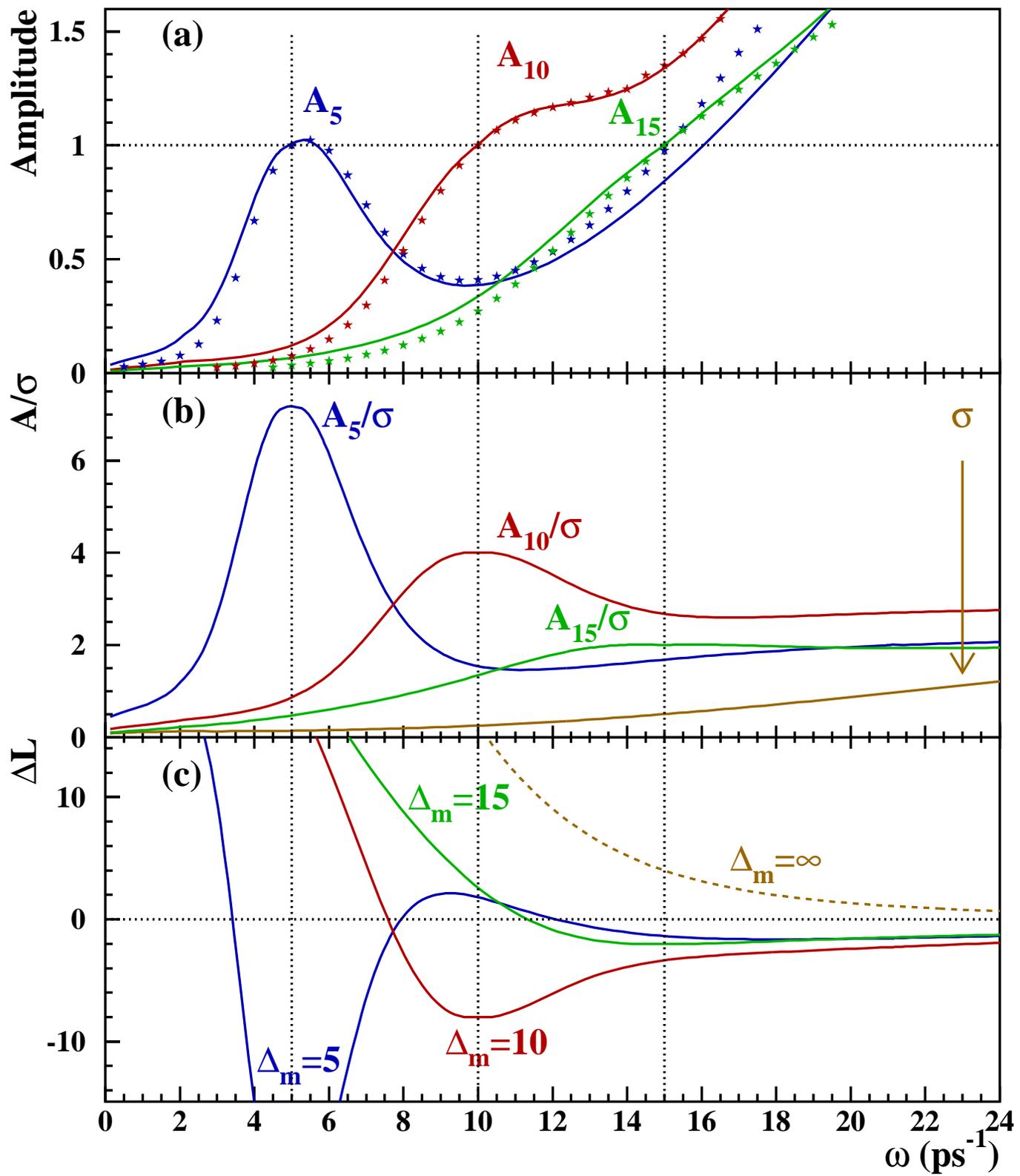
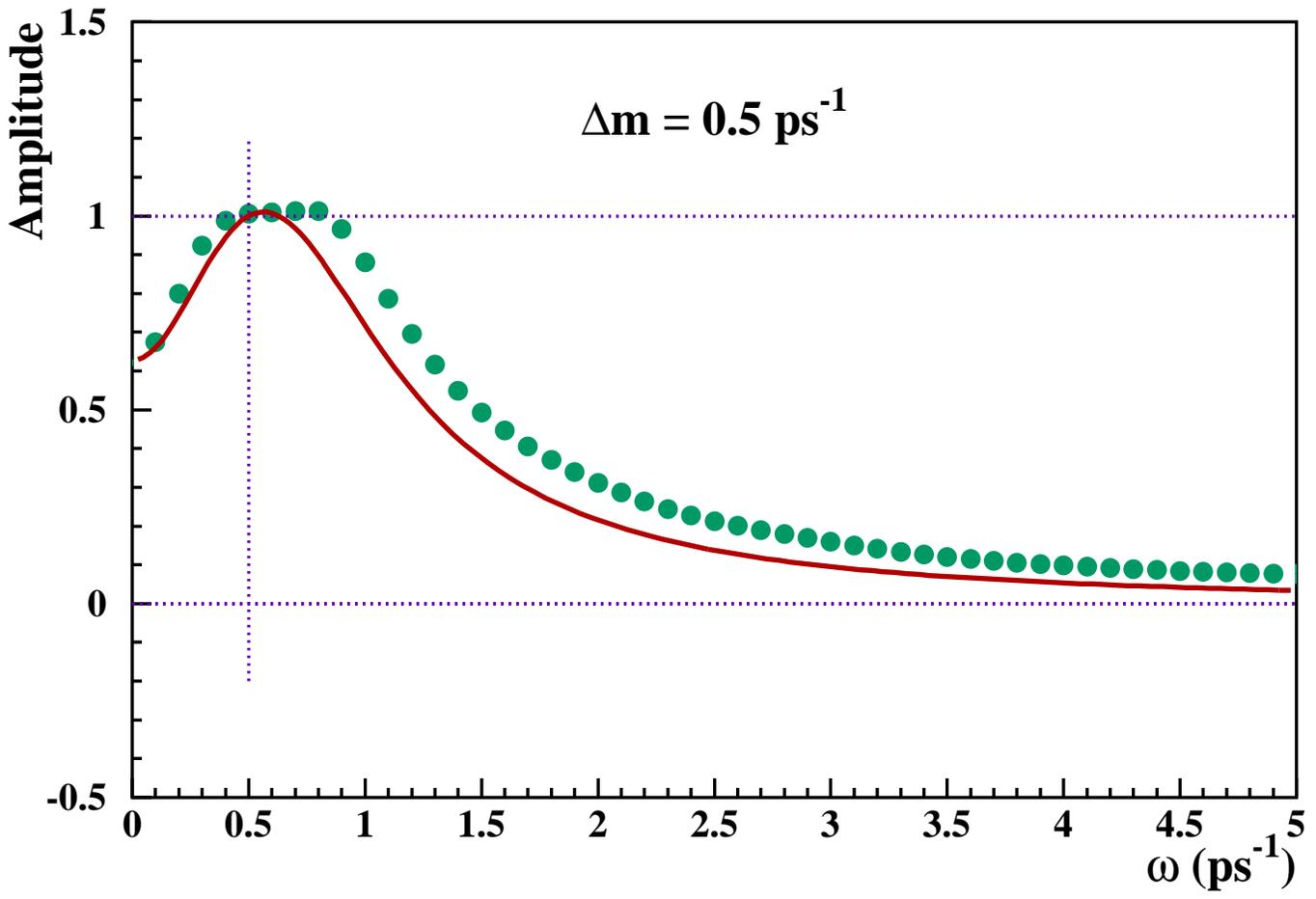
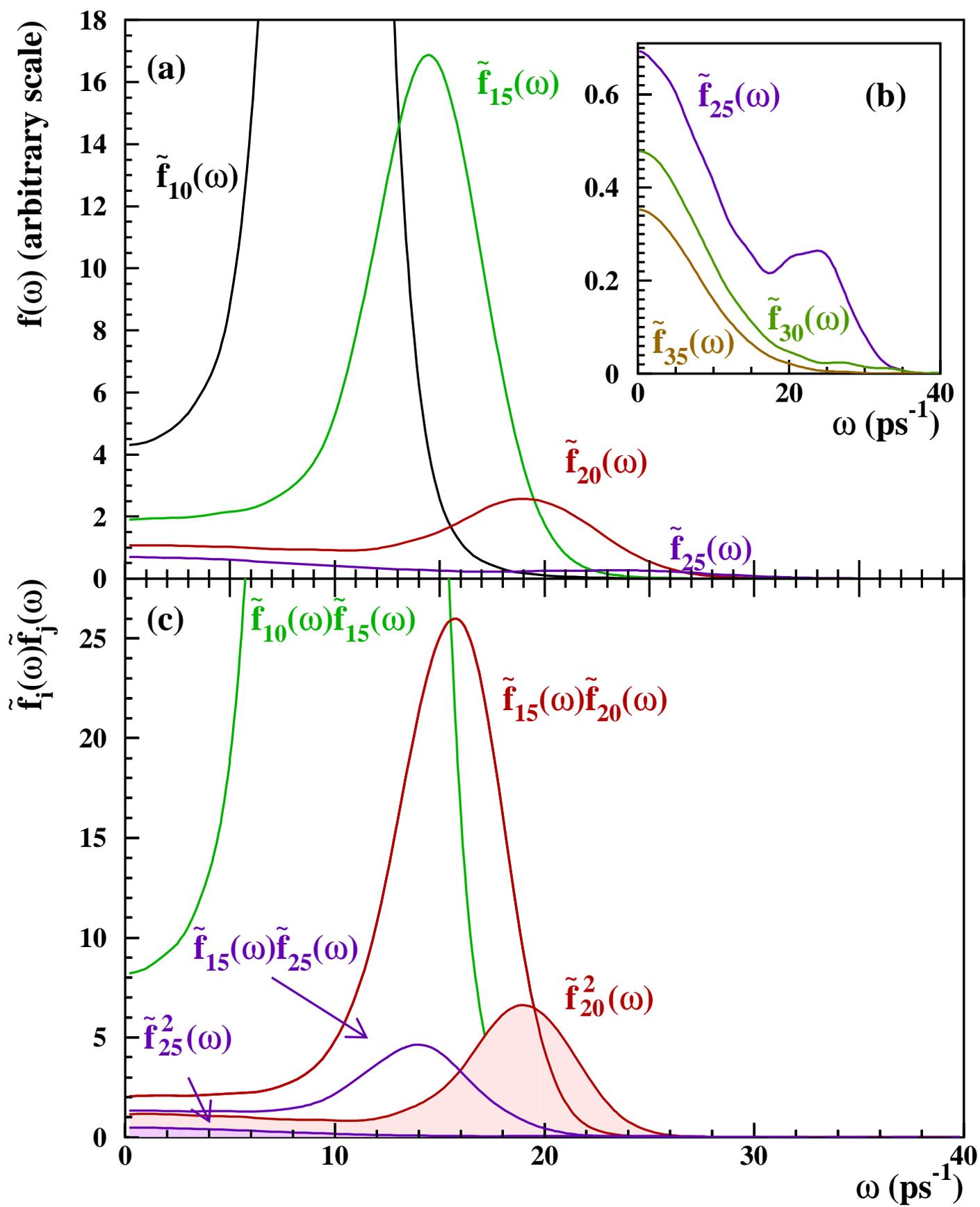
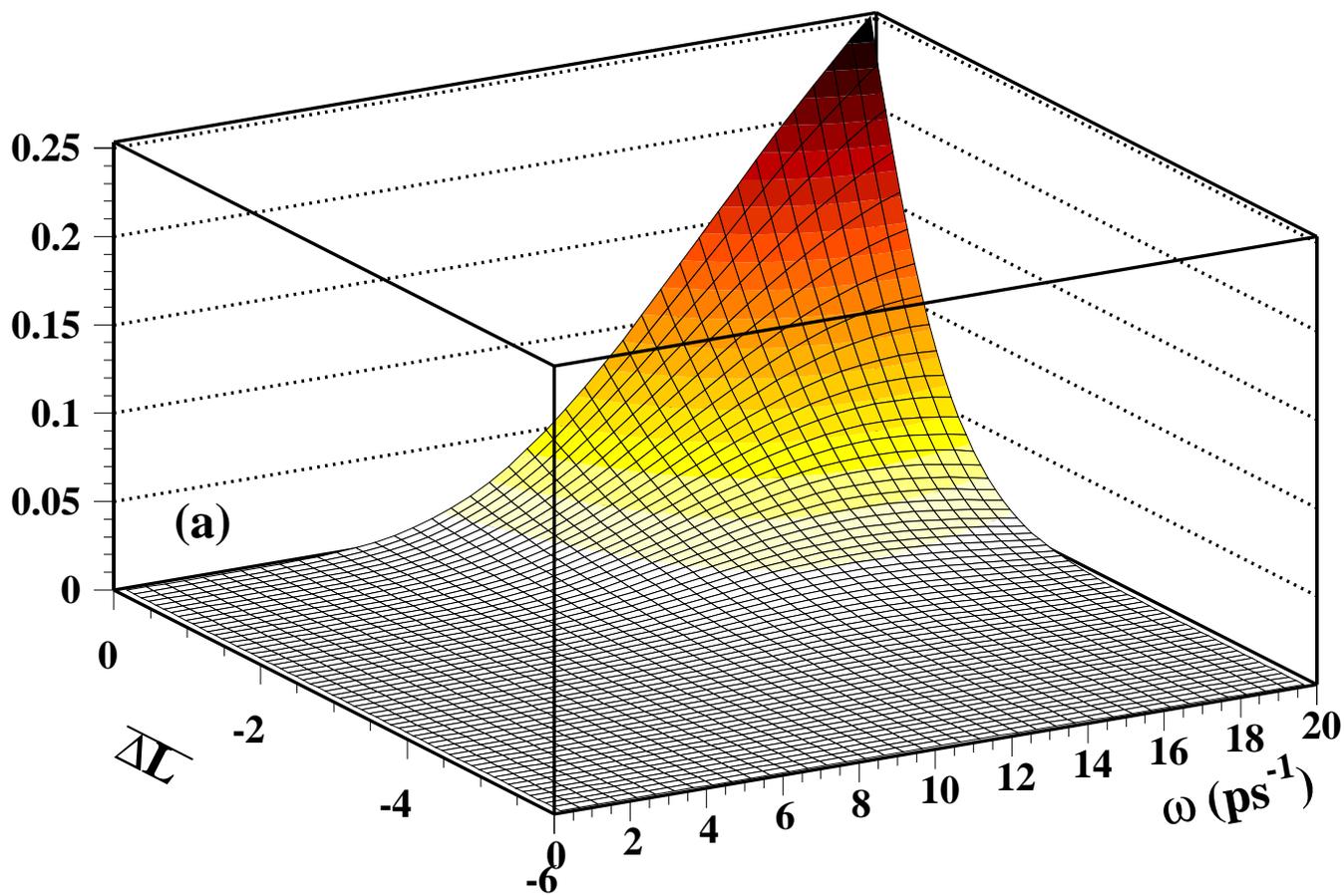
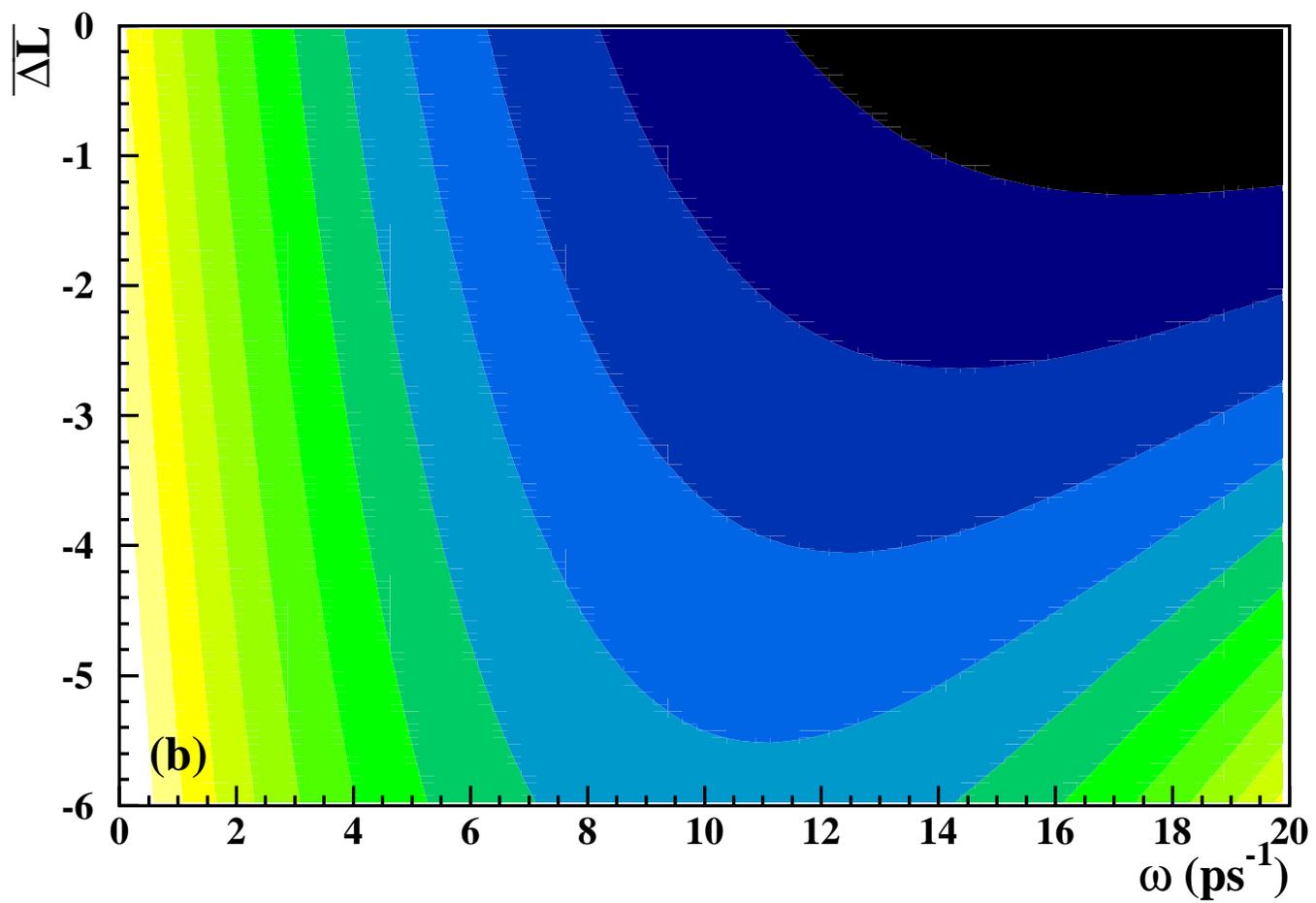
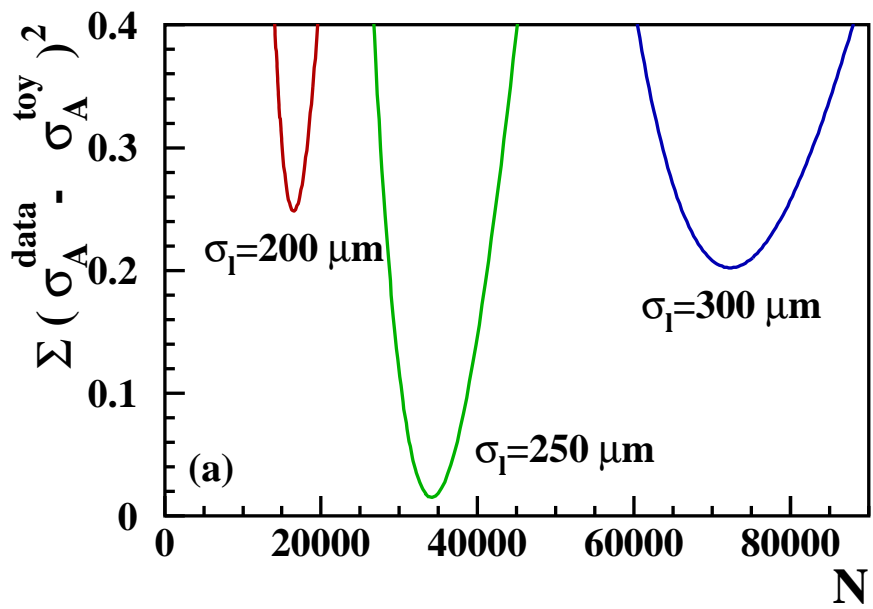
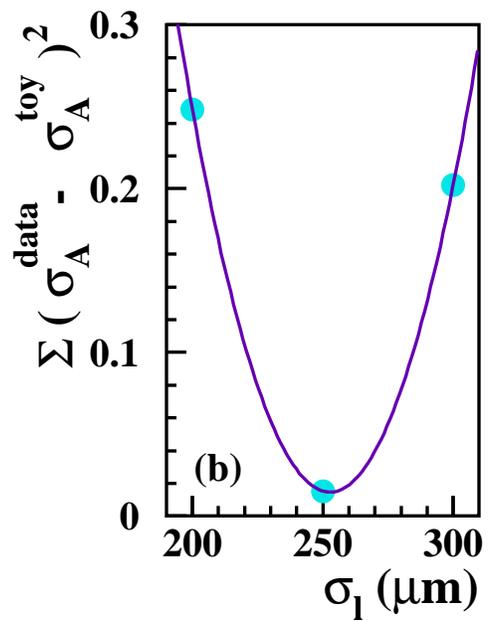
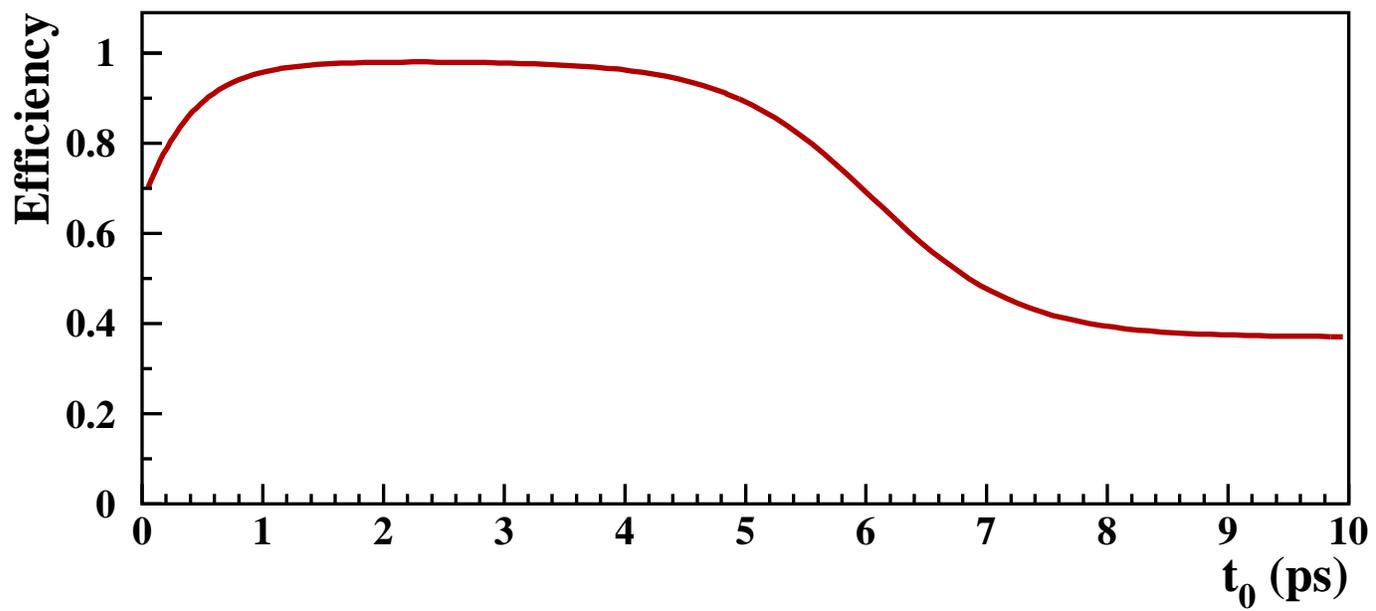
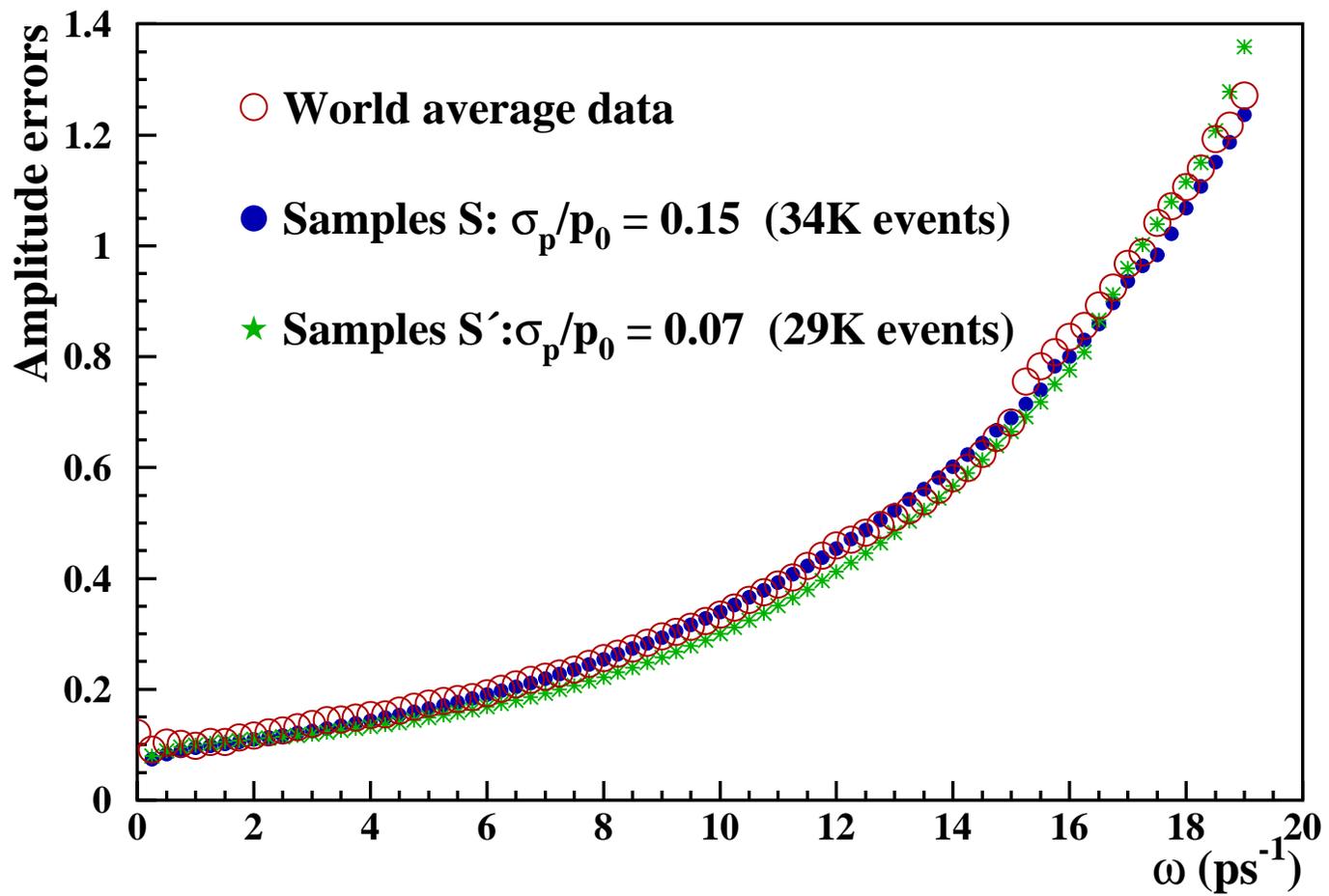
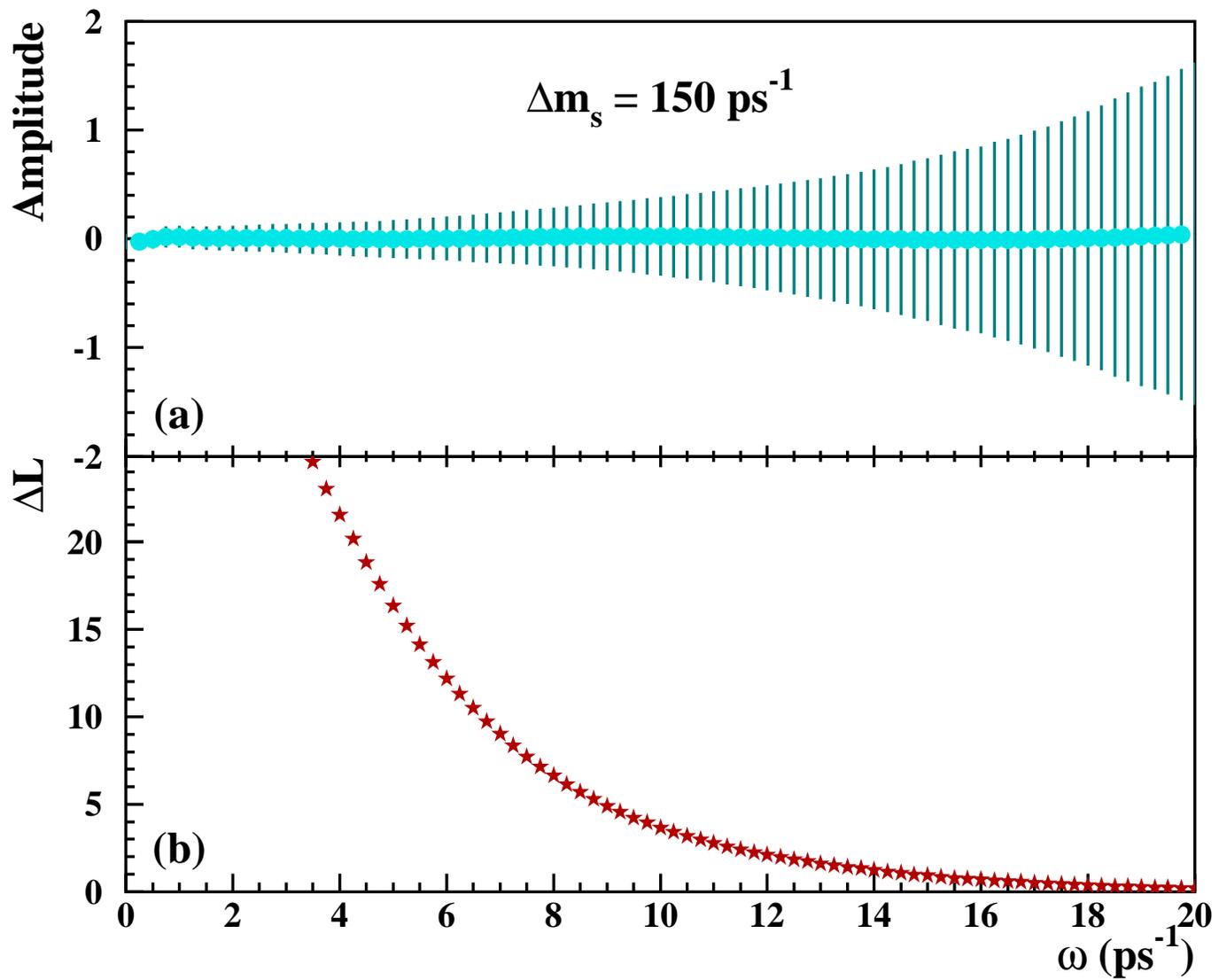
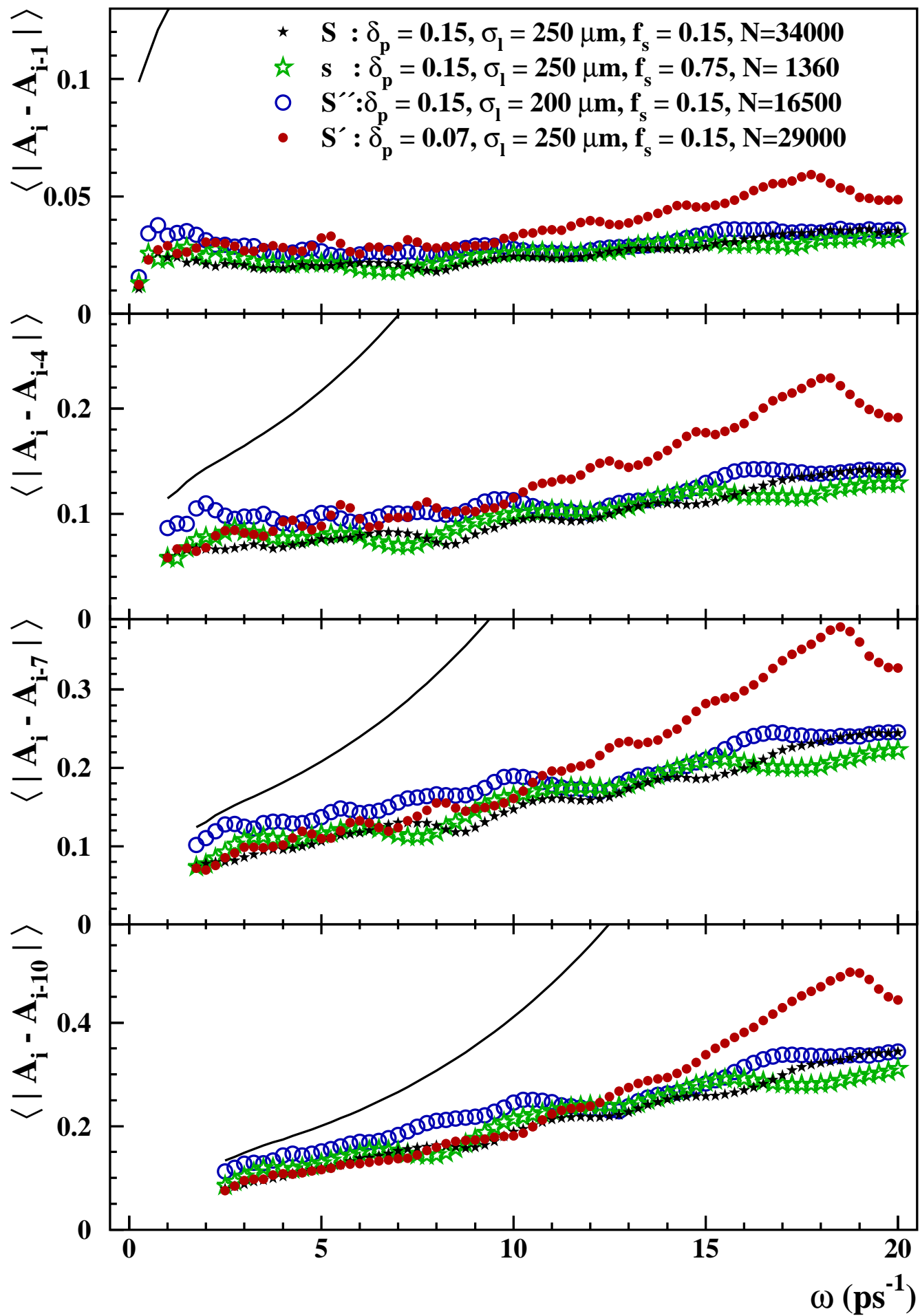
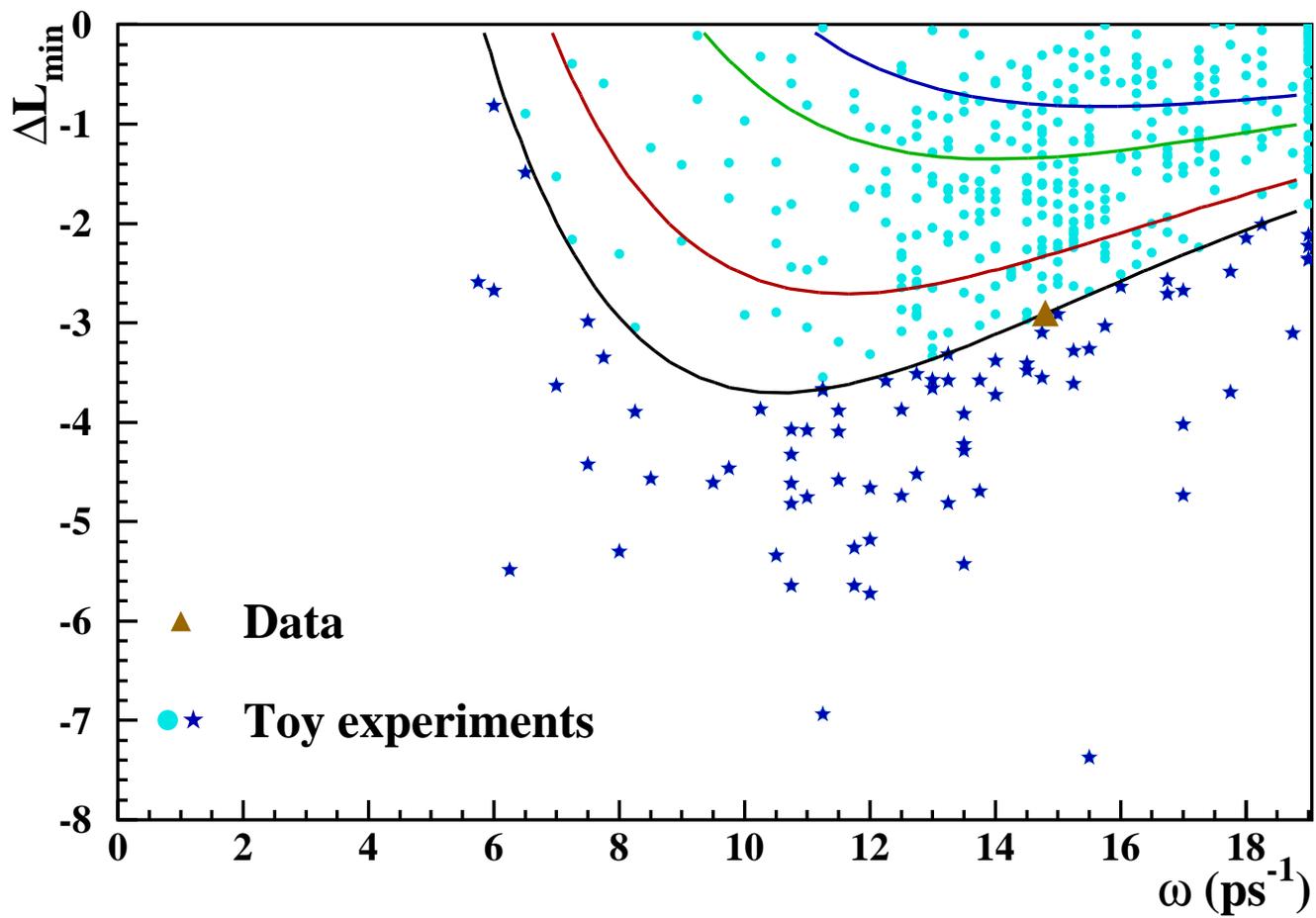
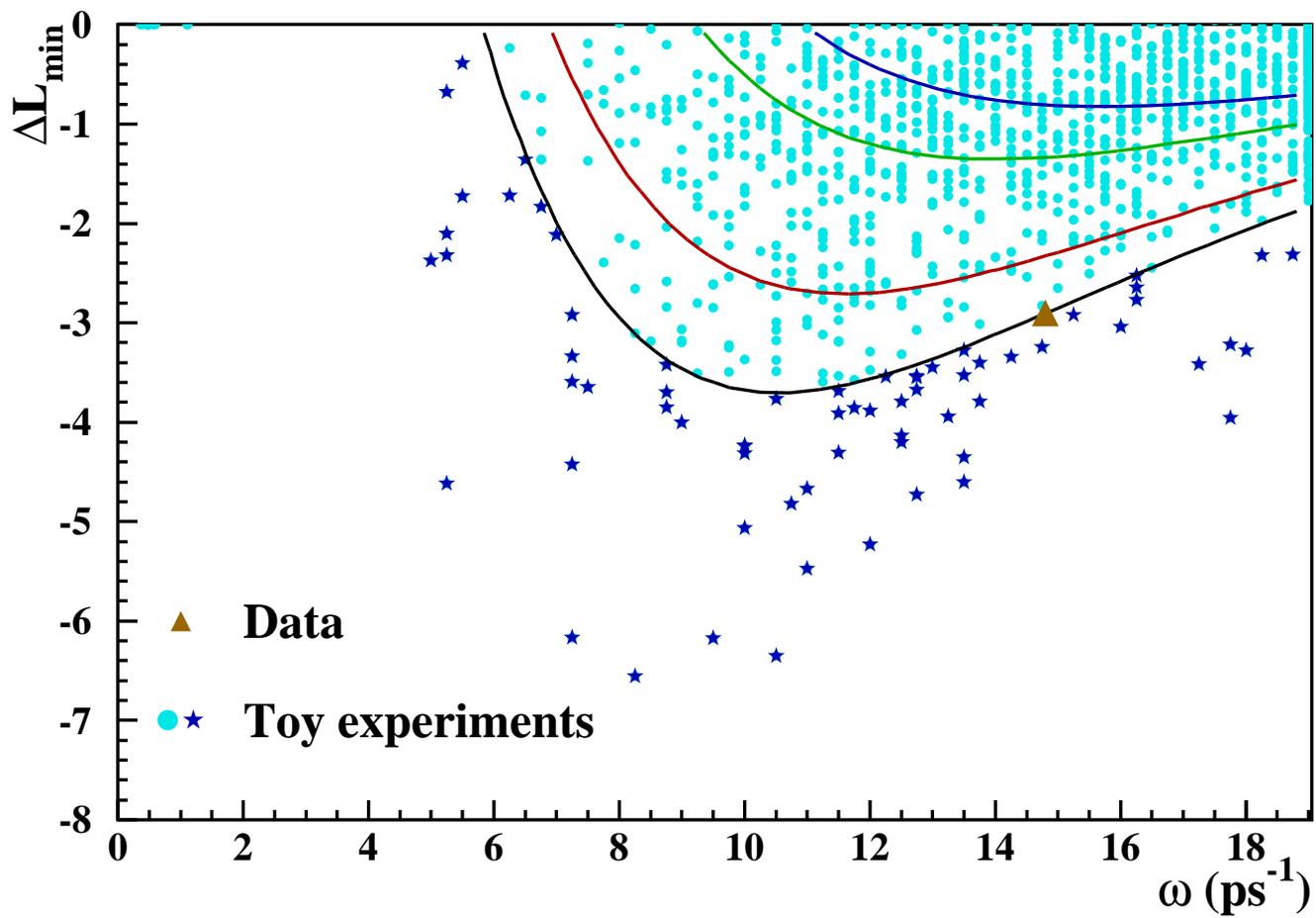
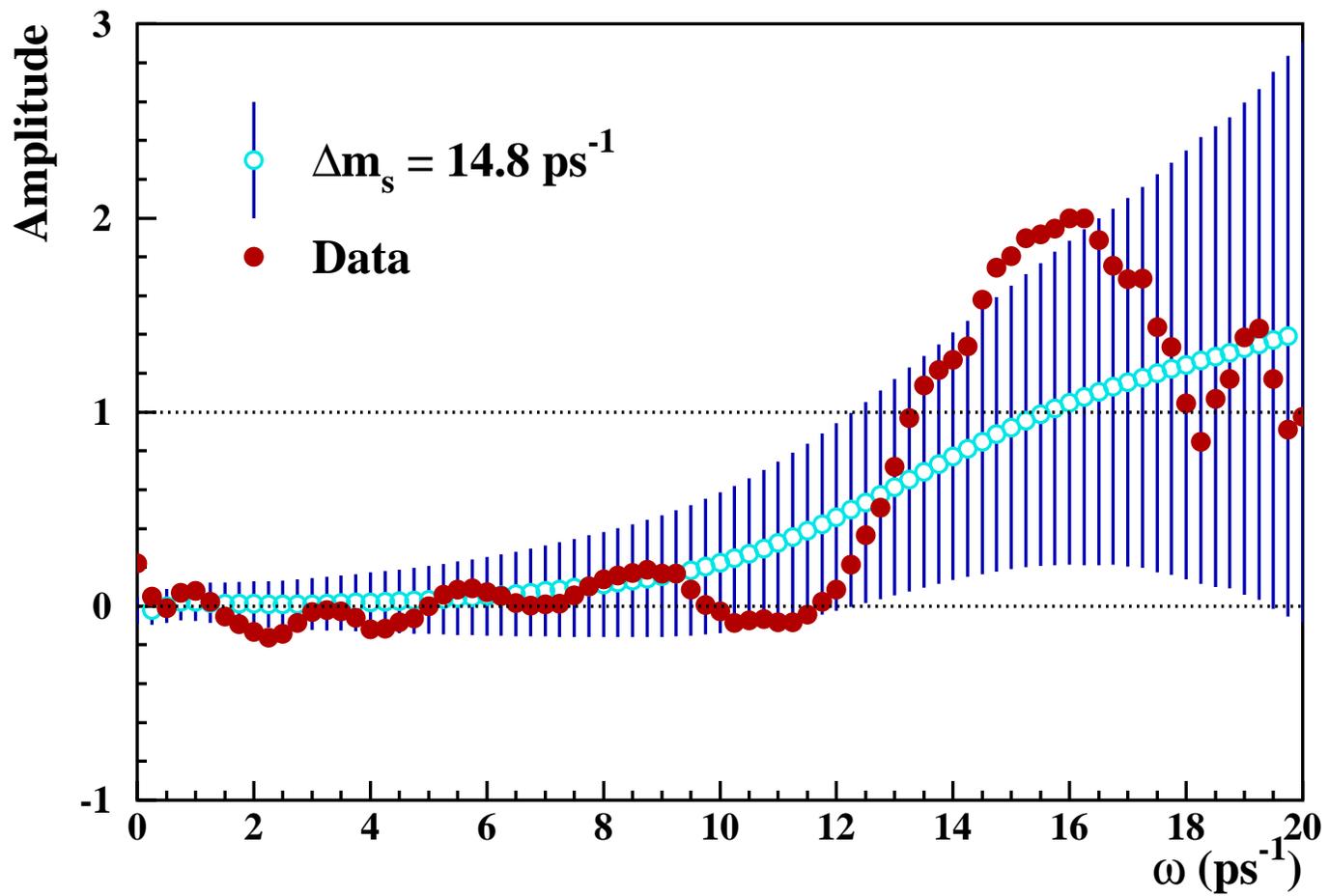
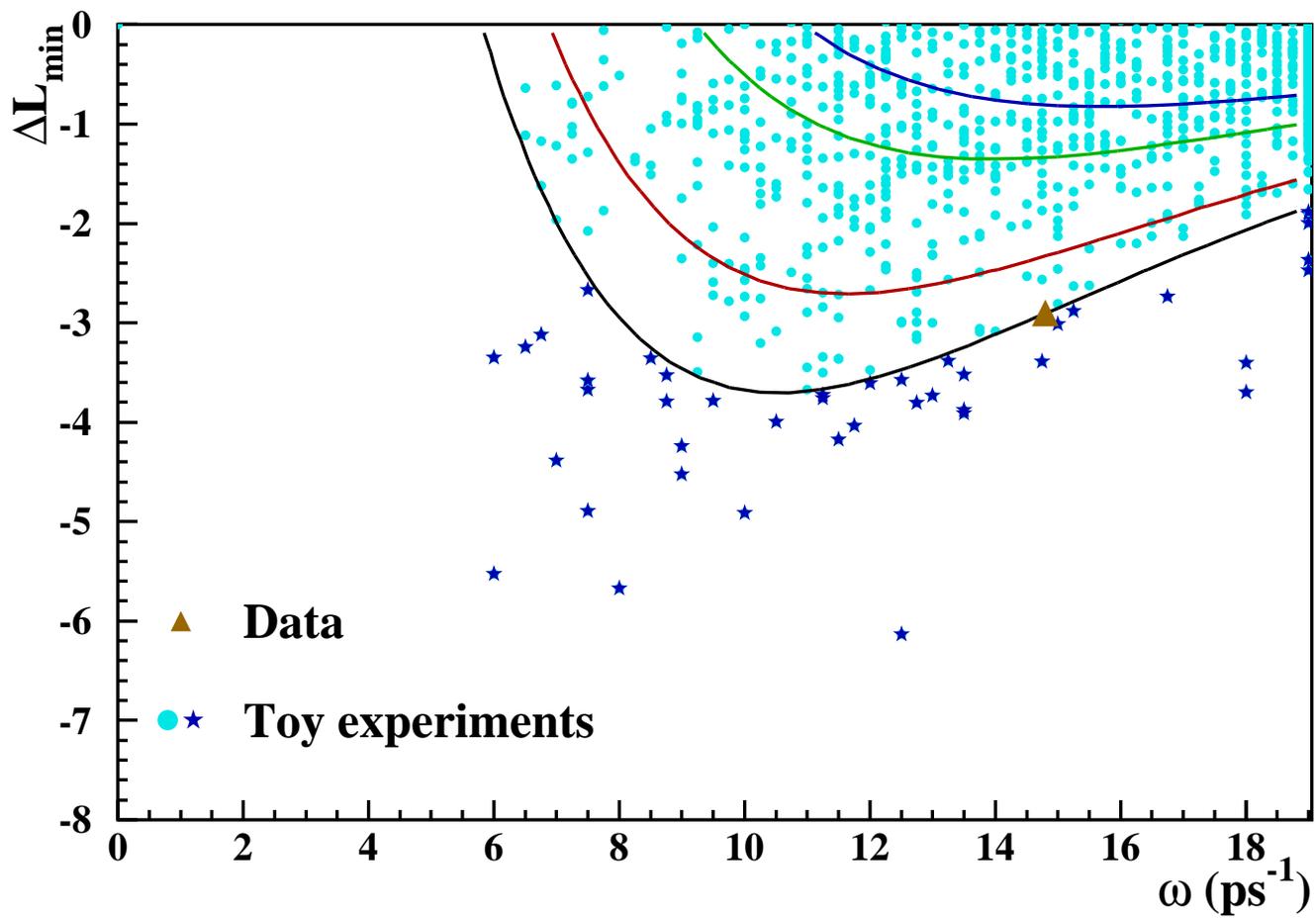
\end{document}